\documentclass[journal]{IEEEtran}
\IEEEoverridecommandlockouts
\usepackage{color}
\usepackage{tikz}
\usetikzlibrary{shapes,arrows,arrows.meta}
\usepackage{mathtools}
\usepackage{amssymb}
\usepackage{lipsum}
\usepackage{float}
\usepackage{makecell}
\usepackage{longtable}
\usepackage{stfloats}
\usepackage{multirow}
\usepackage{amsmath}
\usepackage{bm}
\usepackage{algorithm}
\usepackage{algpseudocode}
\usepackage{booktabs}
\usepackage[numbers,sort&compress]{natbib}
\usepackage{balance}
\usepackage{soul}
\algdef{SE}[SUBALG]{Indent}{EndIndent}{}{\algorithmicend\ }%
\algtext*{Indent}
\algtext*{EndIndent}
\usepackage{xcolor}
\usepackage{threeparttable}
\usepackage{natbib}
\usepackage[breaklinks,citecolor=red,urlcolor=red,bookmarks=false,hypertexnames=true]{hyperref}
\allowdisplaybreaks
\usepackage{graphicx}
\usepackage{caption}
\usepackage{subfigure}
\begin{document}

\title{GNN-PMB: A Simple but Effective Online 3D Multi-Object Tracker without Bells and Whistles}

\author{\IEEEauthorblockA{
Jianan Liu\IEEEauthorrefmark{1},
Liping Bai\IEEEauthorrefmark{1},
Yuxuan Xia,
Tao Huang,~\IEEEmembership{Senior Member,~IEEE,}\\
Bing Zhu\IEEEauthorrefmark{2},~\IEEEmembership{Member,~IEEE}, and 
Qing-Long Han,~\IEEEmembership{Fellow,~IEEE}
}
\vspace{-5 mm}

\thanks{This paper has been accepted by IEEE Transactions on Intelligent Vehicles, DOI: 10.1109/TIV.2022.3217490.}
\thanks{\IEEEauthorrefmark{1}Both authors contribute equally to the work and are co-first authors.}
\thanks{\IEEEauthorrefmark{2}Corresponding author.}
\thanks{J.~Liu is with Vitalent Consulting, Gothenburg, Sweden. Email: jianan.liu@vitalent.se.}
\thanks{L.~Bai and B.~Zhu are with the School of Automation Science and Electrical Engineering, Beihang University, Beijing 100191, P.R.~China. Email:
bai\_liping@buaa.edu.cn (L. Bai);
zhubing@buaa.edu.cn (B. Zhu).}
\thanks{Y.~Xia is with Department of Electrical Engineering, Chalmers University of Technology, Gothenburg 41296, Sweden. Email: yuxuan.xia@chalmers.se.}
\thanks{T.~Huang is with College of Science and Engineering, James Cook University, Smithfield QLD 4878, Australia. Email: tao.huang1@jcu.edu.au.}
\thanks{Q.-L.~Han is with the School of Science, Computing and Engineering Technologies, Swinburne University of Technology, Melbourne, VIC 3122, Australia. Email: qhan@swin.edu.au.}
}

\markboth{IEEE Transactions on Intelligent Vehicles, Oct~2022}%
{\MakeLowercase{\textit{et al.}}: Demo of IEEEtran.cls for IEEE Journals}

\maketitle


\begin{abstract}

Multi-object tracking (MOT) is among crucial applications in modern advanced driver assistance systems (ADAS) and autonomous driving (AD) systems. The global nearest neighbor (GNN) filter, as the earliest random vector-based Bayesian tracking framework, has been adopted in most of state-of-the-arts trackers 
in the automotive industry. The development of random finite set (RFS) theory facilitates a mathematically rigorous treatment of the MOT problem, and different variants of RFS-based Bayesian filters have then been proposed. However,  their effectiveness in the real
ADAS and AD application is still an open problem. In this paper, it is demonstrated that the latest RFS-based Bayesian tracking framework could be superior to typical random vector-based Bayesian tracking framework 
via a systematic comparative study of both traditional random vector-based Bayesian filters with rule-based heuristic track maintenance and RFS-based Bayesian filters on the nuScenes validation dataset. An RFS-based tracker, namely Poisson multi-Bernoulli filter using the global nearest neighbor (GNN-PMB), is proposed to LiDAR-based MOT tasks. This GNN-PMB tracker is simple to use, and it achieves competitive results on the nuScenes dataset. Specifically, the proposed GNN-PMB tracker outperforms most state-of-the-art LiDAR-only trackers and LiDAR and camera fusion-based trackers, ranking the $3^{rd}$ among all LiDAR-only trackers on nuScenes 3D tracking challenge leader board\footnote{https://bit.ly/3bQJ2CP} at the time of submission. Our code is available at {\url{https://github.com/chisyliu/GnnPmbTracker}}.

\end{abstract}



\IEEEpeerreviewmaketitle
\setcitestyle{square}

\section{Introduction}
\label{introduction}

\IEEEPARstart{M}{ulti}-object tracking (MOT) is an integral and critical computational module for various systems, including autonomous vessels \cite{USV}, roadside traffic monitoring \cite{Roadside_tracking}, advanced driver assistance systems (ADAS)
and autonomous driving (AD) \cite{ADAS}, etc. Motivations of using a multi-object tracker include:
1) the tracker assigns and maintains a unique track ID for the same object throughout the life cycle of a tracking process;
2) the tracker rejects the false detection provided by the object detector;
3) the tracker sustains the tracking process when the tracked object fails to be detected over consecutive frames; and
4) the tracker refines the state information provided by the upstream module, e.g. bounding boxes estimated by object detector using mono-camera \cite{monocamera_object_detection} or LiDAR \cite{CenterPoint}, instance clusters estimated by radar \cite{radar_instance_segmentation_1}\cite{radar_instance_segmentation_2}, etc., to support the downstream module like prediction \cite{Prediction_survey}.
Essentially, a multi-object tracker is a state estimator, or equivalently, a filter. In this paper, the concepts "multi-object tracker," "tracker," and "filter" are equivalent.

\begin{figure}
\begin{center}
\includegraphics[scale=0.56]{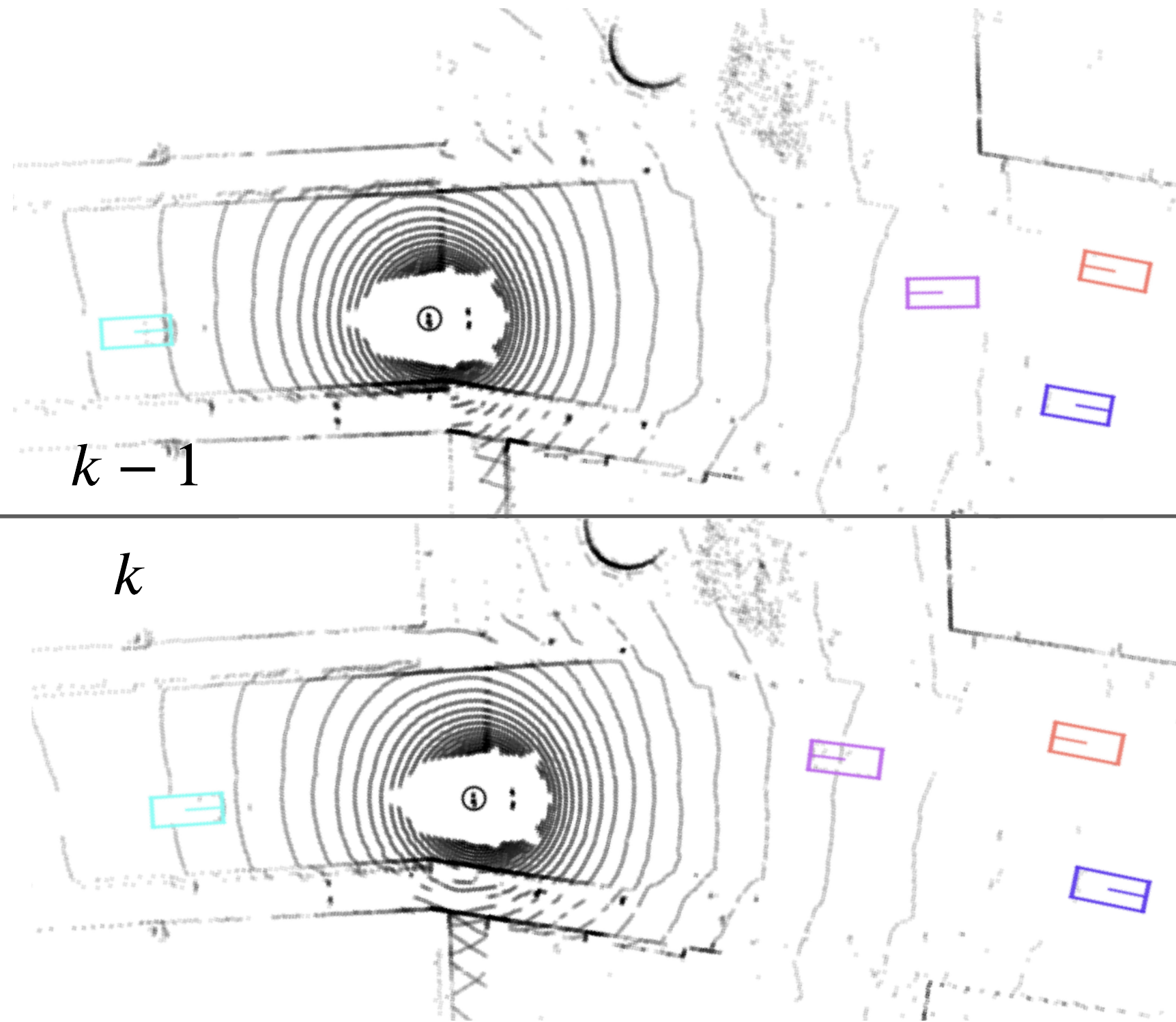}
\caption{Tracking-by-Detection 3D MOT using LiDAR for ADAS and AD: The outputs of the MOT filter over two consecutive frames at time step $k-1$ and $k$ are presented. The ego vehicle is the object surrounded by concentric circles. The LiDAR points are illustrated in the bird-eye-view (BEV) frame using grey dots. The projection of 3D detected bounding boxes onto the BEV frame are represented by colored rectangles, with their orientations indicated by lines. Bounding boxes with the same color share the same track ID.}
\label{fig:fig1}
\end{center}
\end{figure}

Depending on the employed sensor modalities, the task of MOT can be categorized into 2D MOT with camera \cite{zhang2021fairmot}\cite{zhang2021bytetrack}, 3D MOT with LiDAR only \cite{AB3DMOT}\cite{StanfordIPRL-TRI}\cite{SimpleTrack}, and 3D MOT with both camera and LiDAR \cite{Probabilistic3DMM}\cite{EagerMOT}. Base on the processing order of detection and tracking, the task of MOT can be further categorized into tracking-before-detection \cite{davey2007comparison}, joint-detection-and-tracking \cite{lu2020retinatrack}, and tracking-by-detection \cite{henriques2012exploiting}.
In this paper, we focus on the tracking-by-detection 3D MOT with LiDAR for ADAS and AD application, which is the process when outputs of an object detector across frames are refined and connected by their corresponding track IDs, as shown by Fig. \ref{fig:fig1}. Specifically, the object detector first provides the bounding box information, then the multi-object tracker refines the information provided by the object detector and assigns the appropriate ID to the bounding boxes. 

%

%


However, the existing 3D MOT strategies and algorithms in literature are more complicated, requiring either additional data-driven feature extraction modules or heuristic tricks and rules for data association and track maintenance \cite{EagerMOT}\cite{AlphaTrack}\cite{PC3T}. The rules may not be easily generalized in different scenarios or computationally feasible for embedded systems in the real-world. In contrast, this paper proposes a simple but effective online 3D multi-object tracker, namely Poisson multi-Bernoulli filter using the global nearest neighbor (GNN-PMB). The proposed GNN-PMB tracker is designed under random fintie set (RFS) framework without requiring any feature extraction modules, heuristic tricks or additional rules, yet outperforms the state-of-the-art performance on the nuScenes \cite{Caesar2020nuScenesAM} 3D LiDAR-based MOT benchmark dataset. The main contributions of this paper are listed as follows:
\begin{itemize}
\item{ 
A systematic comparative study is presented for the first time by employing several commonly-used Bayesian filters and the proposed GNN-PMB filter, with the different object detectors, on the nuScenes benchmark dataset.
This paper is based on real-world data and provides guidelines for designing the tracking framework in practice.
}
\item{

A simple but effective 3D online LiDAR-based tracker, GNN-PMB, is proposed. The proposed GNN-PMB is designed in a unified and 
simple RFS framework, naturally allowing for a mathematically rigorous treatment of the data association hypothesis and track maintenance. 
It requires no extra feature extraction module, heuristic tricks, or rules for data association and track maintenance. 
Therefore, it is simple to tune and use, 
and more robust to detection errors than the random vector-based GNN framework with many heuristics, which has been widely applied in the practical ADAS and AD system. 
}
\item{
The performance of the proposed tracker is evaluated on the nuScenes dataset, the results show that the proposed LiDAR only-based tracker outperforms the vast majority of the state-of-the-art LiDAR-only trackers, and it is even superior to many state-of-the-art trackers based on fusion of LiDAR and camera for the 3D MOT task.
}
\end{itemize}

The rest of the paper is arranged as follows. Section \ref{relatedwork} discusses the related works of 3D LiDAR-based MOT in autonomous driving-related applications. Section \ref{background} introduces the important RFSs, the modeling assumptions, and the RFS Bayesian recursion. A brief overview and systematic study of typical Bayesian MOT filters are presented and discussed in Section \ref{systematicstudy}. The proposed GNN-PMB tracker is detailed in Section \ref{proposed}. Experimental results are provided and analyzed in Section \ref{result}. Finally, the conclusion is drawn in Section \ref{conclusion}.

\section{Related Works}\label{relatedwork}

%

In this paper, we focus on the task of 3D MOT with LiDAR only, which can be further differentiated into MOT without deep learning and MOT with deep learning. 
In this section, we briefly review these approaches.
In addition, we also review MOT methods with LiDAR and Camera fusion.%

\subsection{3D MOT with LiDAR Only}

\subsubsection{LiDAR-based 3D MOT without Deep Learning}

Most current trackers for 3D MOT tasks in the ADAS and AD system utilize random vector-based GNN frameworks with many tricks and heuristics for track-and-detection data association and track maintenance. AB3DMOT \cite{AB3DMOT} establishes a baseline for LiDAR-based 3D MOT problem. In this algorithm, the track-and-detection association is computed based on the 3D intersection over union (IoU) score \cite{zheng2020diou}, and the track-and-detection association with the highest association score is regarded as the valid association scheme. Probabilistic 3D MOT \cite{StanfordIPRL-TRI} which achieves better results than AB3DMOT, was the first work applying Mahalanobis distance instead of 3D IoU to evaluate track-and-detection association in the 3D MOT problem. SimpleTrack \cite{SimpleTrack} uses a generalized 3D intersection over union (GIoU) instead of 3D IoU as the track-and-detection association score, and it uses a non-maximum suppression (NMS) to process detection information provided by the object detector. This work has demonstrated that combining GIoU and NMS preprocessing improves the tracking result. ACK3DMOT \cite{ACK3DMOT} proposes a joint probability function of appearance, geometry, and distance correlation among the detected bounding boxes and predicted objects to form the cost matrix for track-and-detection data association, leadin to improved tracking performance when combined with an adaptive cubature Kalman filter.

To improve track continuity, both ImmortalTracker \cite{ImmortalTracker} and PC3T \cite{PC3T} evaluate the similar idea of silently maintaining the tracks even when the tracks are no longer visible, which help reduce the ID switches and fragmented tracks. Score refinement is proposed in confidence-based 3D MOT \cite{CBMOT} for track maintenance, where the tracker achieves a low ID switch score and track fragmentation score.

There are also LiDAR only 3D MOT trackers that do not use GNN for data association. RFS-M3 \cite{Pang20213DMT} firstly applies a RFS-based method, specifically the PMBM filter, to the LiDAR 3D MOT problem, where the data association problem is addressed by propagating multiple data association hypotheses with the highest likelihoods over time. 
Belief Propagation Tracker \cite{BPTracker}, was the first time formulated the MOT problem in its factor graph representation. Then, the sum-product message passing algorithm is used to compute the approximate marginal association probability iteratively.


\subsubsection{LiDAR-based 3D MOT with Deep Learning}

SimTrack \cite{SimTrack} introduces an end-to-end trainable model for joint detection and tracking from raw point clouds. A graph structure based on neural message passing was designed in OGR3MOT \cite{OGR3MOT} to process detection and tracking in an online manner, 
where the data association is fully trainable. Neural enhanced belief propagation (NEBP) tracker \cite{NEBP} leverages the belief propagation tracker with a learned graph neural network, and it achieves the state-of-the-art performance on 3D LiDAR MOT.

\subsection{3D MOT with LiDAR and Camera Fusion}

In principle, the performance of MOT with LiDAR only input can be improved by considering LiDAR and Camera fusion \cite{Probabilistic3DMM}\cite{EagerMOT}\cite{AlphaTrack}. The Probabilistic 3D Multi-Modal MOT fuses features from 2D images and 3D LiDAR point clouds to capture the appearance and geometric information of objects \cite{Probabilistic3DMM}. In addition, a new metric that combines the Mahalanobis and feature distances is used for the track-and-detection association. This results show that, by incorporating the camera information, the Probabilistic 3D Multi-Modal MOT tracker achieves better tracking results than the Probabilistic 3D multi-object tracker \cite{StanfordIPRL-TRI} that only uses 3D LiDAR detection as input. Similar to DeepFusionMOT \cite{DeepFusionMOT}, EagerMOT \cite{EagerMOT} increases the tracking performance, compare to Probabilistic 3D Multi-Modal MOT, by utilizing the two-stage data association scheme for 3D detection fused from both LiDAR and camera and 2D detection from the camera, respectively. Later, AlphaTrack \cite{AlphaTrack} further improves the performance of EagerMOT by adding a feature extractor along with the object detector. The feature extractor takes image information and the LiDAR point cloud information as inputs and provides information for the track-and-detection association.

\section{Basis of RFS-based Method}
\label{background}

\subsection{Set Notation}

In RFS-based methods, object states (e.g., position, orientation, size of bounding box, etc.) and their corresponding measurements are represented in the form of finite sets.
It is assumed that there are $n_k$ objects at time $k$. Let $x^i_k$ denotes the state of $i$-th object at time $k$. Thus, the multi-object state at time $k$ can be represented as a finite set $X_k = \{x^1_k, \dots, x^i_k, \dots, x^{n_k}_k\}$.
The cardinality of this set is $|X_k| = n_k$.
In addition, assume there are $p_k$ measurements at time $k$. Let $Z_k = \{z_1,...,z_{p_k}\}$ denote the set of measurements at time step $k$ and let $Z^k = \{Z_1, \dots, Z_k\}$ denotes the sequence of all the measurement sets received so far up to and including time step $k$. 

\subsection{Key Random Processes}
\label{sec_key_random_processes}

Two random processes have prominent roles in the RFS-based methods: Poisson point process (PPP) and Bernoulli process. 
The PPP is a random set of points where the cardinality of the set is Poisson distributed. Similarly, the Bernoulli process is a random set of points where the cardinality of the set is distributed based on the Bernoulli process. 
%
%

A PPP can be described using its intensity function $\lambda(\cdot)$, and in this way its RFS density is
\begin{equation}
    f^{\mathrm{ppp}}(X) = e^{-\int \lambda(x) dx}\prod_{x\in X}\lambda(x),
\end{equation}
%
%
where $X$ is a finite set and its cardinality $|X|$ is Poisson distributed with mean $\bar{\lambda} = -\int \lambda(x) dx$. 

A Bernoulli process with existence probability $r$ and existence-conditioned probability density function (PDF) $f(\cdot)$ has RFS density
\begin{equation} 
f^{\mathrm{ber}}({X}) = 
      \begin{cases} 
             1-r     &  X=\emptyset;     \\ 
             r f(x)  &  X=\{x\};         \\ 
             0       &  \mathrm{otherwise}.
      \end{cases}
\label{eq_SingleBerProcess}
\end{equation}
The cardinality $|X|$ is Bernoulli distributed with parameter $r$. 
%
%
\emph{It is worth mentioning that the existence probability is used for the proposed GNN-PMB tracker to indicate the likelihood that the object exists at the current time step.}

A  multi-Bernoulli (MB) RFS $X$ is the union of a finite number of independent Bernoulli processes $X_1, \dots, X_n$, and its density is 

\begin{equation}
    f^{\mathrm{mb}}(X) = \sum_{\uplus_{i=1}^n X_i= X}f_i^{\mathrm{ber}}(X_i)
\end{equation}
where $\uplus$ denotes the disjoint union and $f_i^{\mathrm{ber}}(\cdot)$ is the density of the $i$-th Bernoulli component as shown in (\ref{eq_SingleBerProcess}).

\subsection{Bayesian Recursion and Multi-object Models}
\label{sec_multiobject_models}

In Bayesian MOT filtering, the multi-object posterior $f_{k|k}(X_k|Z^k)$ is critical because it captures all the information of the set $X_k$ of object states at time step $k$ conditioned on all the measurements. 
The posterior is computed by recursively applying the Chapman-Kolmogorov prediction
\begin{multline}
\label{eq_prediction}
f_{k|k-1}(X_{k}|Z^{k-1}) \\= \int \Phi_{k|k-1}(X_{k}|X_{k-1})f_{k-1|k-1}(X_{k-1}|Z^{k-1})dX_{k-1}
\end{multline}
and the Bayes update 
\begin{equation}
    f_{k|k}(X_k|Z^k) \propto G(Z_k|X_k)f_{k|k-1}(X_k|Z^{k-1}),
\end{equation}
where $\int f(X) d X$ is the set integral  \cite{Mahler2007StatisticalMI}, $\Phi_{k|k-1}(X_{k}|X_{k-1})$ is the multi-object transition density for modeling the dynamics of multiple objects, and $G(Z_k|X_k)$ is the multi-object measurement likelihood for modeling the measurement of multiple objects. In this paper, the standard multi-object dynamic model $\Phi_{k|k-1}(X_{k}|X_{k-1})$ is based on the following assumptions \cite{Mahler2007StatisticalMI}:

\begin{itemize}

    \item Single object with state $x_{k-1}$ at time $k-1$ moves to a new state $x_{k}$ with a Markov transition density $\phi(x_{k}|x_{k-1})$.

    \item Single object with state $x_k$ at time step $k$ has a probability $1 - P^\mathrm{S}(x_k)$ of leaving the sensor's field-of-view, where the superscript $\mathrm{S}$ refers to survival.

    \item The state of newborn objects $X^{\mathrm{b}}_{k}$ at time $k$ appears in sensor's field-of-view according to a PPP with intensity $\lambda^{\mathrm{b}}_{k}(\cdot)$.

    \item The appearing/disappearing of newborn/existing objects and the object motions are conditionally independent of the previous multi-object state $X_{k-1}$.

    \item The set $X_{k}$ of objects at time $k$ is the union of the set $X^{\mathrm{e}}_{k}$ of existing objects at time $k$ and the set $X^{\mathrm{b}}_{k}$ of newborn objects, i.e., $X_{k} = X^{\mathrm{e}}_{k} \cup X^{\mathrm{b}}_{k}$, where the superscript $\mathrm{e}$ refers to existing, and superscript $\mathrm{b}$ refers to newborn.

\end{itemize}

\emph{Specifically, the multi-object transition density from time ${k-1}$ to ${k}$, for the proposed GNN-PMB tracker is given by a convolution of a PPP density for newborn objects $X^\mathrm{b}_{k}$ at time ${k}$ and a multi-Bernoulli density for existing objects $X^\mathrm{e}_{k}$ inherited from time ${k-1}$} as:
\begin{multline}
    \Phi_{k|k-1}(X_{k}|X_{k-1})\\ = \sum_{X_{k}=X^\mathrm{b}_{k} \uplus X^\mathrm{e}_{k} }f_{k}^\mathrm{ppp}(X^b_{k})f_{k|k-1}^\mathrm{mb}(X^\mathrm{e}_{k}|X_{k-1}).
\end{multline}

The standard multi-object measurement model $G(Z_k|X_k)$ is made with the following assumptions:

\begin{itemize}

    \item The measurement set $Z_k$ at time step $k$ consists of measurements $Z^\mathrm{o}_k$ generated by the state set $X_k$ of objects and clutter measurements $Z^\mathrm{c}_k$, i.e., $Z_k = Z^\mathrm{o}_k \cup Z^\mathrm{c}_k$, where the superscript $\mathrm{o}$ refers to the set generated by the set of objects, and the supercript $\mathrm{c}$ refers to the set generated by the clutter process.

    \item 
    $Z^\mathrm{o}_k$ and $Z^\mathrm{c}_k$ are statistically independent.

    \item No measurement is generated by more than one object.

    \item Given a state set $X_k$ of objects, each object state $x_k \in X_k$ is either detected with probability $P^\mathrm{D}(x_k)$, where the superscript D stand for detection probability, and generates a single measurement $z_k$ with measurement likelihood $g(z_k|x_k)$, or misdetected with probability $1-P^\mathrm{D}(x_k)$.

    \item 
    $Z^\mathrm{c}_k$ follows a PPP with intensity $\lambda^\mathrm{c}_k(\cdot)$.

\end{itemize}

\emph{Specifically for GNN-PMB tracker, the multi-object measurement likelihood $G(Z_k|X_k)$ is given by a convolution of a PPP density for clutter measurements $Z^c_k$ and a multi-Bernoulli density for object-oriented measurements $Z^o_k$} by
\begin{equation}
    G(Z_k|X_k) = \sum_{ Z_k = Z^\mathrm{c}_{k} \uplus Z^\mathrm{o}_{k}} f_k^\mathrm{ppp}(Z^\mathrm{c}_{k})f_{k}^\mathrm{mb}(Z^\mathrm{o}_{k}|X_k).
\end{equation}

\section{A Systematic Study of Bayesian MOT Methods}
\label{systematicstudy}

In this section, we present a systematic comparative study of common Bayesian MOT methods from mainly two aspects: 1) track maintenance and 2) approximation methods for computational tractability.

\subsection{Global and Local Hypotheses}

The main challenge of MOT is the unknown data association due to the unknown correspondence between objects and measurements. 
Therefore we start by giving a unified terminology of data association hypotheses.
Consider the data association at time step $k$. A \emph{local hypothesis} is defined as a pair of the object-to-measurement association at time $k$, and a \emph{global hypothesis} is a valid collection of local hypotheses, explaining the association of every object and measurement at time $k$.

\begin{figure}[ht]
  \centering
  \includegraphics[scale=0.27]{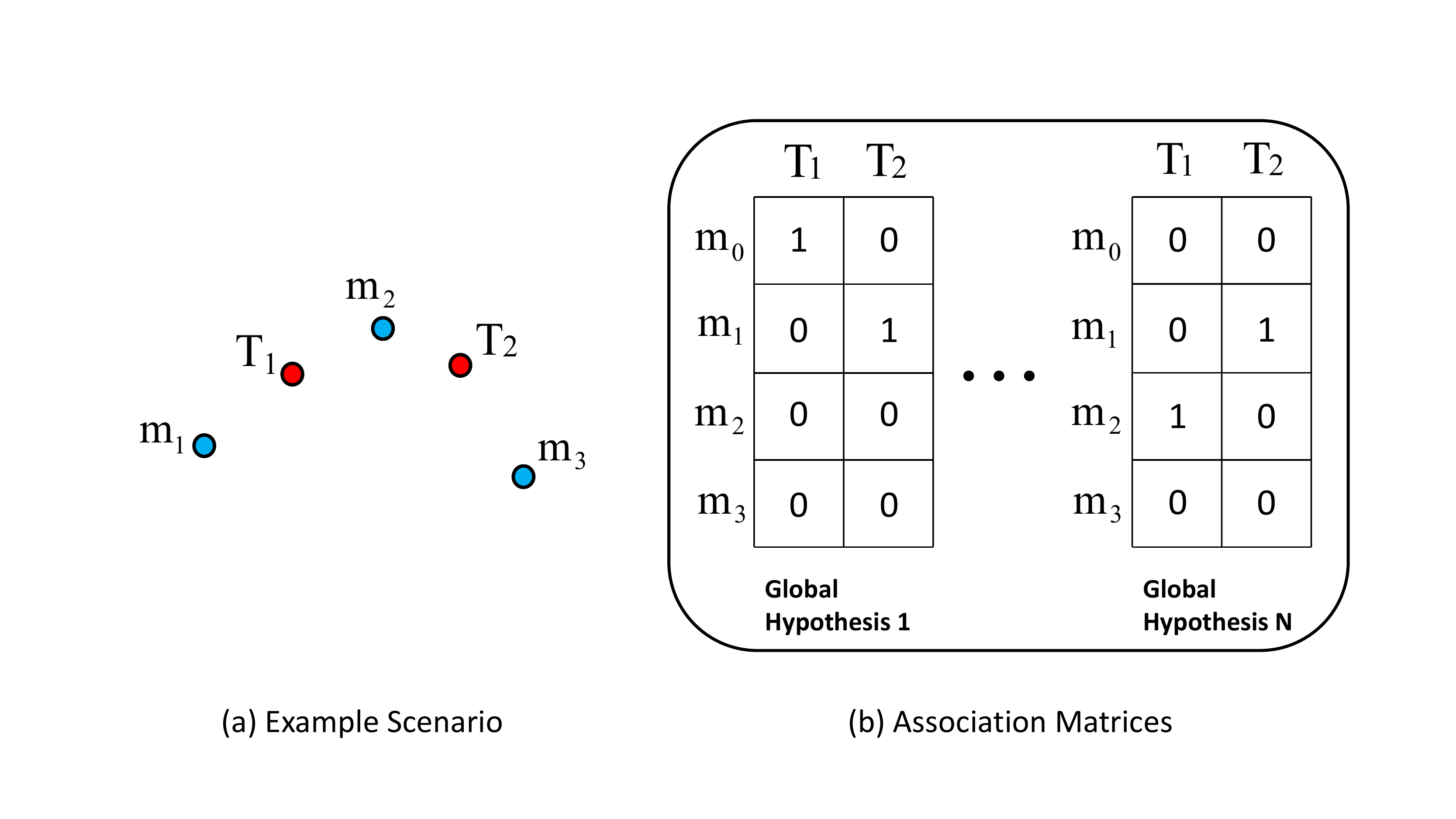}  
  \caption{An example illustrating the local and global hypotheses. Given the objects (denoted by $\text{T}$) and measurements (denoted by $\text{m}$) in (a), examples of global hypotheses, represented using association matrices, are shown in (b). The $1$ and $0$ in the association matrix indicates whether the local hypothesis exists: $0$ means the measurement is not associated with the target, and $1$ otherwise. 
  }
  \label{fig_example}
\end{figure}

To further elaborate, let us consider the example illustrated in Fig. \ref{fig_example} (a) where there are two objects $\text{T}_1, \text{T}_2$, and three measurements $\text{m}_1,\text{m}_2,\text{m}_3$.
 The global hypotheses describing their associations are represented using binary association matrices where each entry represents a possible local hypothesis, and we use dummy notation $\text{m}_0$ for misdetections. 
Each association matrix must satisfy: 1) each column must sum to one, and 2) each row must sum to one or zero. 
An all-zero row, except for $\text{m}_0$, means that the corresponding measurement belongs to the clutter. 
For the two global hypotheses shown in Fig. \ref{fig_example} (b), Global Hypothesis $1$ represents that object $\text{T}_1$ is misdetected and object $\text{T}_2$ is associated with measurement $\text{m}_1$, whereas Global Hypothesis $N$ represents that object $\text{T}_1$ is associated to measurement $\text{m}_2$ and object $\text{T}_2$ is associated to measurement $\text{m}_1$.

\subsection{Different Bayesian MOT Methods}

Bayesian MOT methods can be generally categorized into 1) MOT methods based on random vectors and 2) MOT methods based on RFSs. 

\subsubsection{Vector-based MOT Methods}

Vector-based MOT methods describe the multi-object states and measurements by random vectors, and the most representative methods are
%
%
GNN filter \cite{blackman1999design}, joint probabilistic
data association (JPDA) filter \cite{fortmann1983sonar}\cite{BarShalom2009ThePD}, and multiple hypothesis tracking (MHT) filter \cite{Reid1978AnAF}. 

Due to the unknown data associations, the number of global hypotheses increases hyper-exponentially over time. The GNN filter only keeps the most likely global hypothesis at each time step. 
The JPDA filter first computes the (approximate) marginal object-to-measurement association probabilities and then merges local hypotheses corresponding to the same object. 
MHT seeks to find the most likely global hypothesis over a sliding window of consecutive time steps, which involves the propagation of multiple global hypotheses over time.

\subsubsection{Set-based MOT Methods}

Set-based MOT methods describe the multi-object states and measurements by RFSs. There are lots of RFSs-based MOT methods in the literature. 
Early developments include methods that avoid explicitly handling the data association uncertainty, such as probability hypothesis density (PHD) filter \cite{mahler2003multitarget} and cardinalized PHD (CPHD) filter \cite{mahler2007cphd} using moment approximations. 

The PHD and CPHD filters approximate the multi-object posterior by a PPP and an i.i.d. cluster process, respectively, both in the sense of minimizing the Kullback-Leibler divergence. The CPHD filter is computationally heavier than the PHD filter, but it yields better performance when the signal-to-noise ratio is low.

A significant trend in RFSs-based MOT is the development of filters based on multi-object conjugate priors, which means that the multi-object posterior has the same functional form as the predicted distribution (and the prior). 
A typical example is the Poisson multi-Bernoulli mixture (PMBM) filter \cite{GarcaFernndez2018PMBM}, which gives the close-form solution for the standard multi-object models introduced in Section \ref{sec_multiobject_models}. In the PMBM filter, both the prediction and the update preserve the PMBM form of the density without approximation:
\begin{align}
    f^\mathrm{pmbm}_{k|k^\prime}(X_{k}|Z^{k^\prime}) &= \sum_{X^u_{k} \uplus X^d_{k} = X_{k}} f^\mathrm{ppp}_{k|k^\prime}(X^u_{k})f^\mathrm{mbm}_{k|k^\prime}(X^d_{k}) \label{eq_pmbm},\\
    f^\mathrm{mbm}_{k|k^\prime}(X^d_{k}) &= \sum_{h=1}^{H_{k^\prime}}w_{k^\prime}^h f^{h}_{k|k^\prime}(X^d_{k}), \label{eq_mbm}
\end{align}
where $k^\prime \in \{k-1,k\}$. 
In \eqref{eq_pmbm}, the set $X^u_{k^\prime}$ of undetected objects that have not yet been detected is described by a PPP, whereas the set $X^d_{k^\prime}$ of detected objects that have been detected at least once is described by a multi-Bernoulli mixture (MBM). In the MBM \eqref{eq_mbm}, each multi-Bernoulli component corresponds to a unique global hypothesis for the detected objects. The $h$-th multi-Bernoulli component has density $f^{h}_{k|k^\prime}(\cdot)$ and weight $w^h$, which satisfies that $\sum_{h=1}^{H_{k^\prime}} w_{k^\prime}^h = 1$. If there is only a single multi-Bernoulli component in \eqref{eq_mbm}, i.e., when $H_{k^\prime}=1$, then the PMBM filter reduces to a PMB filter \cite{williams2015marginal}.

\subsection{Track Maintenance}

For practical implementations of Bayesian MOT methods, an efficient track maintenance scheme is required for tracking a time-varying number of objects. 
In this paper, a \emph{track} is defined as a sequence of local hypothesis densities at consecutive time steps that correspond to the same object. Track maintenance refers to the process where a track is initiated, sustained, and terminated. 
In what follows, we discuss the track maintenance scheme for vector-based and set-based MOT methods separately. 

\subsubsection{Vector-based MOT Methods}

Vector-based MOT methods can maintain track continuity by associating an object state estimate with a previous state estimate. 
However, they mainly rely on heuristic methods to consider the appearance/disappearance of new/existing objects. 
A commonly used rule for track maintenance is called $M$/$N$ logic.
Specifically, a tentative track is initiated when a measurement is not associated with existing tracks. 
This tentative track is confirmed if there are $M$ measurement associations out of $N$ consecutive time steps. The termination of tracks follows a similar procedure. 
Alternatively, one can use track-score-based logic for track maintenance by performing hypothesis tests.

\subsubsection{Set-based MOT Methods}

The RFS formalism facilitates modeling the appearance/disappearance of new/existing objects in a Bayesian setting. 
For example, in the PMBM and PMB filters, we only extract object state estimates from Bernoulli components with existence probability above a certain threshold. 
However, in RFS-based MOT methods, time sequences of tracks cannot be constructed easily as the multi-object states are order-independent. One approach to maintaining track continuity is to add unique labels to the object states and form tracks by linking object state estimates with the same label \cite{aoki2016labeling}. 
A more appealing approach to solve the track-building problem is by computing multi-object densities on sets of trajectories \cite{garcia2019trajectory}, and a typical example is trajectory PMBM filter \cite{granstrom2018poisson}. 
We note that the prediction and update in the PMBM filter can be seen as an efficient method for calculating the time marginals of the RFS of trajectories \cite{granstrom2018poisson}. 
Therefore, track continuity in the PMBM and PMB filters is implicitly maintained and can be established using metadata.
The detailed procedure is described in Section \ref{TrackMaintenance}.

\subsection{Approximations for Computational Tractability}

Practical MOT implementations need efficient approximations to keep the computational complexity at a tractable level. 
The approximations methods can be categorized into local and global hypothesis reductions.

\subsubsection{Local Hypothesis Reduction}

The commonly used strategy to limit the number of local hypotheses is gating. Specifically, for a predicted local hypothesis density, only the associations of measurements inside its gated region are considered to reduce the number of updated local hypotheses. 
In addition, different MOT methods use diverse techniques to reduce the number of local hypotheses further after the update. 
For example, in MHT, the number of local hypotheses is limited by implementing $N$-scan pruning \cite{Blackman2004MultipleHT} or by pruning local hypotheses with low scores. 
In Gaussian implementations of PHD and CPHD filters, Gaussian mixture reduction is performed for the PPP intensity by pruning components with small weights and merging similar components. 
In PMBM and PMB filters, it is necessary to prune Gaussian components in the PPP intensity with small weights and Bernoulli components with small existence probabilities.

\subsubsection{Global Hypothesis Reduction}

The key to global hypothesis reduction relies on how to solve the data association problem efficiently. 
For cracking the multi-scan data association problem, typical solutions include Lagrangian relaxation \cite{Blackman2004MultipleHT} and Markov chain Monte Carlo sampling \cite{Oh2009MarkovCM}. 
For MOT methods considering the single-scan data association problem, the most likely global hypothesis can be obtained by solving a 2D assignment problem using algorithms such as the Hungarian algorithm \cite{crouse2016implementing}.
The $H_{k^\prime}$ best global hypotheses can be obtained using Murty's algorithm \cite{crouse2016implementing}. 
The merging step in the JPDA filter and the track-oriented PMB (TO-PMB) filter uses the (approximate) marginal association probabilities.
They can be either computed using the $H_{k^\prime}$ best global hypotheses obtained from Murty's algorithm or directly obtained using loopy belief propagation (LBP) \cite{williams2014approximateEO} without explicit enumeration of global hypotheses.

\section{The New GNN-PMB Tracker}        
\label{proposed}

According to the analysis presented in \cite{xia2017performance} and \cite{smith2019systematic}, the PMBM filter results in the best performance in simulation, but it is computationally intensive. Thus, PMBM may not be suitable for the real-world scenario due to existence of many objects and the corresponding measurements per time step, as well as the management of massive hypotheses.

There are two common approaches to reducing the computational burden introduced in the hypothesis management part of the PMBM filter. The first approach is to merge different local hypotheses corresponding to the same Bernoulli component, e.g., using LBP, and the resulting filter is called TO-PMB \cite{williams2015marginal}. The second approach is to only propagate the most likely global hypothesis. In this case, the PMBM recursion is reduced to the PMB recursion with much lower computations. In this paper, we adopt the second approach and propose the GNN-PMB tracker for LiDAR-based 3D MOT. 

\begin{figure*}
    \centering
    \includegraphics[scale=0.57]{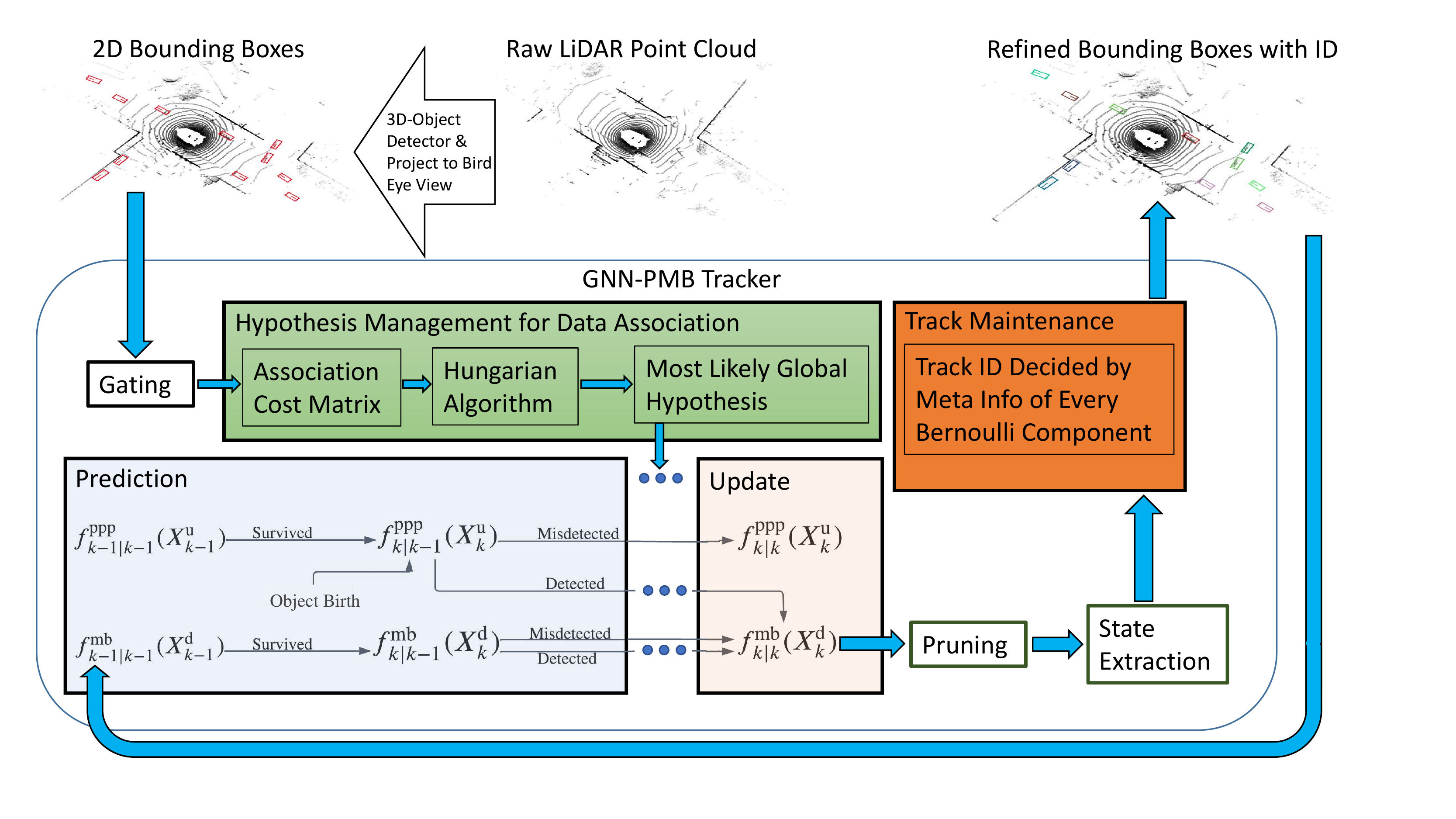}
    \caption{The illustration of entire framework of the proposed GNN-PMB tracker, which consists of the recursion of PPP and the recursion of MB, hypothesis management, pruning, state extraction, and track maintenance.}
    \label{gnn_pmb_tracker_pipeline}
\end{figure*}

In this section, we first explain the overall framework of the proposed GNN-PMB tracker, as illustrated in Fig. \ref{gnn_pmb_tracker_pipeline}. Then, we elaborate
on the details of two core modules, hypothesis management and
track maintenance. At last, we discuss the rest of the modules and their corresponding parameters.
%

\subsection{The Framework of GNN-PMB Tracker}
The available information of the 3D bounding boxes provided by the LiDAR object detector includes (x, y, z) coordinate in the global frame, bounding box size, orientation, velocity, detection score, and class type information. However, to keep the multi-object tracker as simple as possible, the proposed tracker only requires the $x$ and $y$ coordinates in the global Cartesian frame as input. Other information provided by the 3D LiDAR object detector is propagated without filtering. As shown in Fig. \ref{gnn_pmb_tracker_pipeline}, the 2D bounding boxes in BEV, as the measurements of possible objects and clutters, would be utilized as input to the tracker. After the gating, the most likely global hypothesis is decided in hypothesis management procedure, by applying Hungarian algorithm to the association cost matrix, formed using input 2D bounding boxes and the predicted PPP and MB component. Then the updated local hypothesis densities are obtained by standard Kalman update according to the most likely global hypothesis. At last, pruning and state extraction from the MB are implemented afterward to refine the object states. In the last, the ID information of every object is given according to the meta information attached to every Bernoulli component to generate the trajectories.

\subsection{Hypothesis Management and Track Maintenance for GNN-PMB Tracker}

\subsubsection{Hypothesis Management}
\label{HypothesisManagement}

\begin{figure*}[ht]
\begin{center}
\includegraphics[scale=0.5]{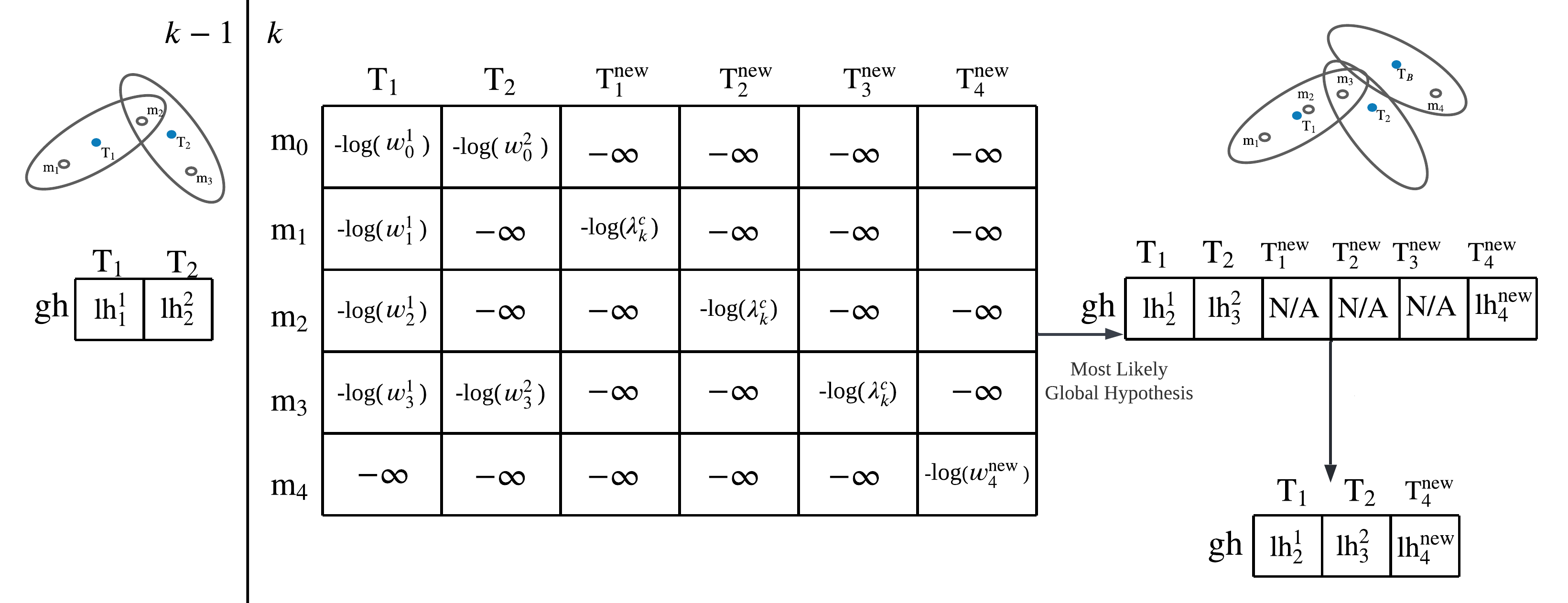}
\caption{The hypothesis management procedure of GNN-PMB tracker. 
A cost matrix is generated based on the most likely global hypothesis at time $k-1$ and then propagated to the current time $k$. 
At time $k$, the cost matrix is organized and the most likely global hypothesis 
is selected by solving a 2D assignment problem using this cost matrix. 
In this figure,  $lh^t_m$ denotes the local hypothesis with measurement $m$ and object $t$, $gh$ stands for global hypothesis, $\lambda^c_k(\cdot)$ is the clutter intensity, and $w$ is the likelihood score of the local hypothesis.
}\label{gnnpmb_hypothesis}
\end{center}
\end{figure*}

The hypothesis management procedure, as illustrated in Fig. \ref{gnnpmb_hypothesis}, is one of the core modules of the proposed GNN-PMB tracker. At time $k-1$, the selected most likely global hypothesis consists of two local hypotheses, as indicated by the left side of the figure. The local hypothesis $lh_1^1$ indicates that track $\text{T}_1$ is associated with measurement $\text{m}_1$, and the other local hypothesis $lh^2_2$ stands for the association between track $\text{T}_2$ and measurement $\text{m}_2$. At time $k$, a cost matrix is generated based on the most likely global hypothesis at time $k-1$ as prior information. Let $c_{ij}$ denote the $i$-th row $j$-th column element in the cost matrix, given by the corresponding negative logarithm of the association likelihood $w$. Such cost element for local hypothesis within the gated area of a track is set to be the negative logarithm of 
$w$. For the detected track, $w$ denotes the occurrence likelihood of a local hypothesis:

\begin{equation}
  w_{m_i}^{T_j} =
     \frac{1}{2 \pi} |{ \bf P}|^{-\frac{1}{2}} \exp^ 
           { -\frac{1}{2} ({m_i} - {T_j})^\mathrm{T} { \bf P}^{-1} ({m_i} - {T_j}) },
\end{equation}
where $m_i$ is the 2D position of the $i$-th measurement, $T_j$ is the 2D position of the $j$-th track, and $\bf P$ is the covariance matrix of the Gaussian density. For instance, as indicated by the right side of Fig. \ref{gnnpmb_hypothesis}, the association likelihood score of the local hypothesis where $\text{T}_1$ is associated with $\text{m}_1$ is $w^1_1$.  
Therefore, the element of the cost matrix in the corresponding grid is $c_{11}$=$-\log(w_1^1)$. The costs of local hypotheses  for existing tracks outside the gated area are set to be negative infinity. For instance, the $4^{th}$ measurement lies outside the gating region of the $1^{st}$ track, thus the corresponding cost is negative infinity. Besides, the $\text{m}_0$ in Fig. \ref{gnnpmb_hypothesis} denotes the missed detection hypothesis, which represents the hypothesis that the track is still alive but misdetected. 
Furthermore, any measurement can also be generated by a new track, so $\text{T}_i^{\text{new}}$ is used to indicate the initial detection of a new track. For each measurement, if it is outside the gated area of the Poisson intensity of undetected objects, represented using a Gaussian mixture, the likelihood score of the corresponding new track is given by $\lambda^\mathrm{c}_k(\cdot)$, which is the intensity of PPP represents the clutter measurement set $Z^\mathrm{c}_k$. 
As each measurement initiates a single new track, the likelihood that measurement $\text{m}_i$ creates new track $\text{T}_i^{new}$, where $i \neq i^{new}$ is zero, and therefore its corresponding element in the cost matrix is infinity.

Every feasible global hypothesis at time step $k$ can be represented by a binary association matrix $\text{\textbf{A}}$. Finding the association matrix, that corresponds to the most likely global hypothesis, can be further formulated as a 2D assignment problem in the following form:

\begin{equation}
\begin{aligned}
a_{ij}^\ast= & \arg \min_{a_{ij}}  \sum_{i,j} \|c_{ij}a_{ij}\|\\
\textrm{s.t.} ~~~~ & \sum_i a_{ij}=1,~~
  \sum_j a_{ij}=1,\\
  &a_{ij} \in \{0,1\},
\end{aligned}
\end{equation}
where $c_{ij}$ denotes the cost element in the aforementioned cost matrix; $a_{ij}$ is the selection term of the $i$-th row and $j$-th column of association matrix $\text{\textbf{A}}$. The selection term has a binary value, i.e. it is either $0$ or $1$. $a_{ij}=1$ means the entry at the $i$-th row and $j$-th column is selected, thus the local hypothesis between the corresponding measurement and track is part of the output global hypothesis, and $a_{ij}=0$ means the entry at the $i$-th row and $j$-th column is not selected. This 2D assignment problem can be solved by the Hungarian algorithm.

After obtaining the most likely global hypothesis, the local hypotheses consist of the selected global hypothesis with likelihood scores lower than a predetermined threshold need to be pruned. The final global hypothesis, which is formed by the remaining local hypotheses after pruning, will be propagated to time step $k + 1$.

\subsubsection{Track Maintenance}
\label{TrackMaintenance}
As another critical module of GNN-PMB tracker, track maintenance is achieved by utilizing the meta-data information conveyed in the different Bernoulli components. 
Suppose a measurement is assigned to its corresponding new Bernoulli component instead of an existing one. 
In this case, the new Bernoulli component would be output as a valid object, and the object ID is 
the current maximum ID plus $1$.
The new ID is stored as metadata to the new Bernoulli component.
For the situation that measurement is associated with an existing Bernoulli component in the most likely global hypothesis, the Bernoulli component would be output as an existing object, and the ID of the object remains the same as the one for the existing Bernoulli component.

\subsection{Parameters for the GNN-PMB Tracker}
\label{paradef}


\subsubsection{Gating Distance}

Gating, which prunes away all the detected bounding boxes with a distance to the centroid of predicted tracks smaller than a threshold, before organizing the possible local hypotheses for each track. 
The gating distance can be defined using different metrics, e.g., Euclidean distance, 2D IoU, 2D GIoU, 2D Mahalanobis distance, etc. 
In the proposed GNN-PMB, we choose the 2D Mahalanobis distance which incorporates the uncertainty information as the gating distance, and is easier to compute than the 2D IoU and the 2D GIoU.
It is defined by
\begin{equation}
d=\sqrt{(z_1 - z_2)^\mathrm{T} {\rm \bf P}^{-1} (z_1 - z_2)},
\end{equation}
where $z_1$ and $z_2$ are two points from the same Gaussian distribution with covariance matrix ${\rm \bf P}$. 
%
%

\subsubsection{Clutter Rate}
The clutter generation process is a PPP with intensity $\lambda^c_k(\cdot)$, as discussed in Section \ref{sec_multiobject_models}. 
To keep the computation simple, it is assumed that a constant expected number of clutter is generated uniformly across the field of view (FoV). 
Therefore, the clutter intensity is defined as the expected number of clutters over the area of FoV. 

\subsubsection{Survival Probability}
As specified in Section \ref{sec_multiobject_models}, the survival probability of an object, $P^\mathrm{S}(x_k)$, represents the possibility that an object remains at the next frame. Theoretically, $P^\mathrm{S}(x_k)$ should be defined in terms of the object position $x_k$. E.g., the disappearance of an object probably occurs around the peripheries of the FoV. However, $P^\mathrm{S}(x_k)$ is set to a constant to keep the computation as simple as possible.

\subsubsection{Probability of Detection} 
Also according to Section \ref{sec_multiobject_models}, the probability of detection of a given object $x$ is denoted by $P^\mathrm{D}(x_k)$.
To better capture the likelihood of every detected bounding box in the dynamic environment, $P^\mathrm{D}(x_k)$ is specifically chosen as detection score of each detected bounding box provided by object detector, rather than being set as a constant value. For the Bernoulli component that has previously detected but misdetected at the current time step, the detection score at the previous time step is applied as $P^\mathrm{D}(x_k)$ at the current time step.

\subsubsection{Pruning Threshold of Existence Probability} As explained in \ref{sec_key_random_processes}, a Bernoulli process 
for existing object is characterised by an existence probability $r$ and existence-conditioned PDF $f(\cdot)$.
To reduce computation, all Bernoulli components with existence probability less than the pruning threshold need to be eliminated. To this end, the threshold should be small enough to retain the Bernoulli components for multiple frames before it is discarded.

\subsubsection{Detection Score Threshold} 
The detection score threshold is a predetermined threshold to prune the input of the tracker. 
For instance, a detection score threshold of $0.5$ means only detected bounding boxes with a detection score higher than $0.5$ would be used as input to the MOT tracker. 

\subsubsection{Non-maximum Suppression (NMS) Threshold}
The object detector often creates multiple bounding boxes around the same object, but only one detection is required for each object. 
Non-max suppression is used to suppress the less likely bounding boxes. 
The NMS score is implemented as the 3D-IoU \cite{zheng2020diou} score in GNN-PMB. 
If there are multiple bounding boxes where 3D-IoU score exceeds the specified NMS threshold, only the bounding boxes with the highest detection score would be kept as the tracker input.

\subsubsection{Poisson Birth Density}
The object birth process is also modeled by PPP, as introduced in Section \ref{sec_multiobject_models}. The cardinality is distributed according to a Poisson distribution with intensity $\lambda^b_{k}(\cdot)$. In the proposed GNN-PMB, $\lambda^b_{k}(\cdot)$ is an unnormalized Gaussian mixture with identical weights, and the initial covariance matrix $P_0$ is identical for all the Gaussian components.
Therefore, the weight of the Gaussian distribution and the initial covariance matrix $P_0$ need to be tuned to specify the Poisson birth density.

\subsubsection{State Extraction Threshold} 
As described in \ref{sec_multiobject_models}, the detected multi-object state is modeled by a MB for the proposed GNN-PMB tracker. Each Bernoulli RFS density has its corresponding existence probability. Only the Bernoulli components with an existence probability higher than the specified extraction threshold are considered valid objects for a given frame. Bernoulli RFS density with an existence probability lower than the specified extraction threshold would be silently maintained until the existence probability falls below the pruning threshold.


\section{Experiments and Analysis}
\label{result}

\subsection{Dataset and Evaluation Metrics}
\label{metrics}
There are four major LiDAR-based 3D detection and tracking benchmark datasets, namely KITTI \cite{Geiger2013VisionMR}, Waymo \cite{Sun2020ScalabilityIP}, nuScenes \cite{Caesar2020nuScenesAM}, and Argoverse \cite{Chang2019Argoverse3T}. 
In this work, we select the nuScenes benchmark dataset to conduct experiments as the popular nuScenes dataset provides the most reliable perception situation and a large amount of testing data in diverse scenarios for 3D MOT. 
To properly evaluate our proposed GNN-PMB, we use the following basic evaluation metrics: number of true positive (TP), number of false negative (FN), number of false positive (FP), number of ID switch (IDS), and number of times a trajectory is fragmented (FRAG). In additional, the number of mostly tracked (MT) tracks, which denotes the number of ground-truth trajectories that are tracked for at least $80\%$ of their respective life span, together with the number of mostly lost (ML) tracks, which denotes the number of ground-truth trajectories that are tracked only for at most $20\%$ of their respective life span, are also applied to provide overall evaluation. Moreover, more comprehensive metrics like the average multi-object tracking accuracy (AMOTA) and the corresponding accuracy measurement multi-object tracking precision (AMOTP) are also used in paper. The detailed information of all the metrics mentioned above can be found on the nuScenes 3D MOT challenge website\footnote{https://www.nuScenes.org/tracking}.

\subsection{Comparisons among Different Bayesian Filters}
\label{filtercomp}

In this section, the filter performance in terms of different MOT metrics is discussed, as shown in TABLE \ref{comparison_between_tracker_frameworks_in_nuScenes_validation_dataset}.
All the secondary metrics, including MT, ML, TP, FP, FN, IDS, FRAG, are evaluated under the recall with the highest MOTA.

\begin{table*}[ht]
\caption{Comparison of 
LiDAR-based object detectors and Bayesian multi-object trackers on nuScenes validation set.} 
\label{comparison_between_tracker_frameworks_in_nuScenes_validation_dataset}
\begin{center}
\begin{threeparttable}
\begin{tabular}{ l|l|ccccccccc}
  \hline
  Tracker & Detector & AMOTA$\uparrow$ & AMOTP$\downarrow$ & MT$\uparrow$ & ML$\downarrow$ & TP$\uparrow$ & FP$\downarrow$ & FN$\downarrow$ & IDS$\downarrow$ & FRAG$\downarrow$ \\
  \hline
    & PointPillars & 0.251 & 1.403 & 1924 & 2158 & 50631 & 11776 & 45334 & 5932 & 2579 \\
GNN & Megvii & 0.509 & 0.881 & 3323 & 1722 & 67774 & 10759 & 29703 & 4420 & 1209 \\
    & CenterPoint & 0.603 & 0.735 & 3725 & 1447 & 72192 & 11767 & 25076 & 4629 & 1236 \\
\hline
    & PointPillars & 0.17  &1.472   &1443   &2581   & 35583  & 42811  &52241   &14073   &4728   \\
PHD & Megvii &0.268&1.201   &478   &4670   &17298   &6501   &79652   &4947   &3365    \\
    & CenterPoint &  0.313 & 1.114  &835   &3403   &18856   &39724   &67996   &15045   &6071   \\
\hline
      & PointPillars & 0.185 & 1.380 & 2474 & 2232 & 56685 & 19096 & 44444 & 768 & 857 \\
TO-PMB & Megvii & 0.294 & 0.926 & 2429 & 3162 & 49827 & 18613 & 51722 & 348 & 403 \\
      & CenterPoint & 0.324 & 0.812 & 2737 & 2955 & 53982 & 15382 & 47433 & \textbf{482} & 449 \\
\hline
     & PointPillars & 0.269 & 1.260 & 2709 & 2311 & 59250 & 12060 & 40800 & 1847 & 843 \\
PMBM & Megvii & 0.577 & 0.739 & 4314 & 1528 & 79094 & 13536 & 21829 & 974 & 406 \\
     & CenterPoint & 0.645 & 0.600 & 4591 & 1349 & 82480 & 14608 & 18234 & 1183 & 403 \\
\hline
      & PointPillars & 0.311 & 1.231 & 2754 & 2236 & 60929 & \textbf{9993} & 39945 & 1023 & 769 \\
GNN-PMB & Megvii & 0.619 & 0.716 & 4314 & 1552 & 79434 & 11710 & 21955 & 508 & 372 \\
      & CenterPoint & \textbf{0.707} & \textbf{0.560} & \textbf{4608} & \textbf{1347} & \textbf{83134} & 12362 & \textbf{18113} & 650 & \textbf{345} \\
\hline
\end{tabular}
\begin{tablenotes}
        \footnotesize
        \item[$\uparrow$] The upper arrow indicates that better performance is registered with higher score, same for the following tables. 
        \item[$\downarrow$] The lower arrow indicates that better performance is registered with lower score, same for the following tables.
        \item Note: JPDA and CPHD trackers have also been implemented, but cannot be executed in real-time due that massive number of detection provided by nuScenes dataset strain JPDA and CPHD computation.
\end{tablenotes}
\end{threeparttable}
\end{center}
\end{table*}

\subsubsection{Quality of Detection Input}

%
The tracking performance of a tracker is affected by the performance of the attached detector.
In the experiment, we have used three different object detectors, including the CenterPoint detector, the Megvii detector, and the PointPillar detector, with a descending order of their detection performance in the nuScenes LiDAR-based detection challenge.
The CenterPoint\footnote{https://bit.ly/3bWHSWA} detection result is provided by its authors, and the nuScenes 3D tracking challenge organizer provides the results for PointPillars\footnote{https://www.nuScenes.org/data/detection-pointpillars.zip} and Megvii\footnote{https://www.nuScenes.org/data/detection-megvii.zip}.
As shown in TABLE \ref{comparison_between_tracker_frameworks_in_nuScenes_validation_dataset}, the tracking performances are commensurate to the quality of the detection input for all the evaluated trackers.
The best performance is achieved with the CenterPoint detector input for each tested tracker, whereas the worst performance is achieved with the PointPillars detector.

\begin{table*}[tht]
\caption{Ablation study of filter parameters. This table indicate how the tracking result changes with different parameters.} 
\label{parameters_ablation}
\begin{center}
\begin{threeparttable}
\begin{tabular}{cc|c|cccccccccc}
  \hline
 \multicolumn{2}{c|}{Parameters}        & Value&AMOTA$\uparrow$ & AMOTP$\downarrow$ &RECALL$\uparrow$& MT$\uparrow$ & ML$\downarrow$ & TP$\uparrow$ & FP$\downarrow$ & FN$\downarrow$ & IDS$\downarrow$ & FRAG$\downarrow$ \\
  \hline
    \multicolumn{2}{c|}{\multirow{3}{*}{\begin{tabular}[c]{@{}c@{}}Detection Score\\Threshold\end{tabular}}}&    0.1 & 0.707 & 0.559 & 0.738 & 4606 & 1353 & 83088 & 12300 & 18167 & 642 & 341 \\
       & &0.2 & 0.672 & 0.705 & 0.693 & 4391 & 1414 & 80971 & 9198 & 20318 & 608 & 328 \\
    &    &0.3 & 0.625 & 0.834 & 0.659 & 4266 & 1199 & 80601 & 9965 & 20564 & 732 & 500 \\
 \hline
    \multicolumn{2}{c|}{\multirow{2}{*}{\begin{tabular}[c]{@{}c@{}}NMS Threshold\end{tabular}}}   &0.1 & 0.707 & 0.560 & 0.730 & 4611 & 1329 & 83057 & 12491 & 18082 & 758 & 333 \\
   & &0.98 & 0.680 & 0.572 & 0.715 & 4513 & 1424 & 81830 & 13240 & 19267 & 800 & 333 \\
 \hline
 \multirow{6}{*}{\begin{tabular}[c]{@{}c@{}}Poisson\\Birth\\Density\end{tabular}}&\multirow{4}{*}{\begin{tabular}[c]{@{}c@{}}Weight of\\Gaussian\\Components\end{tabular}}&0.0001     & 0.626 & 0.764 & 0.653  & 4415 & 1588 & 81039 & 14055 & 20254 & 604 & 429 \\
 &&0.001      & 0.702 & 0.555 & 0.729  & 4510 & 1533 & 82235 & 14171 & 19101 & 561 & 351  \\
 &&0.01       & 0.703 & 0.559 & 0.729  & 4618 & 1383 & 83128 & 13714 & 18202 & 567 & 334  \\
&&0.1        & 0.707 & 0.560  & 0.730   & 4611 & 1329 & 83057 & 12491 & 18082 & 758 & 333  \\
\cline{2-13}

        &\multirow{2}{*}{\begin{tabular}[c]{@{}c@{}}Initial\\Covariance\end{tabular}}&15 & 0.699 & 0.566 & 0.730 & 4494 & 1518 & 81942 & 13036 & 19386 & 569 & 291 \\
        &&100 & 0.625 & 0.761 & 0.636 & 4363 & 1646 & 80024 & 13538 & 21169 & 704 & 366 \\
\hline
\multicolumn{2}{c|}{\multirow{3}{*}{\begin{tabular}[c]{@{}c@{}}State Extraction\\ Threshold\end{tabular}}}&0.5    & 0.680  & 0.565 & 0.714  & 4432 & 1539 & 81294 & 12754 & 19993 & 610 & 265  \\
 &&0.7    & 0.688 & 0.561 & 0.728  & 4452 & 1536 & 81625 & 12125 & 19691 & 581 & 287  \\
&&0.9    & 0.698 & 0.565 & 0.733  & 4495 & 1487 & 82057 & 11913 & 19192 & 648 & 295 \\

\hline
\end{tabular}
\end{threeparttable}
\end{center}
\end{table*}

\subsubsection{Performance of GNN, PHD, JPDA, and CPHD}

As shown in TABLE \ref{comparison_between_tracker_frameworks_in_nuScenes_validation_dataset}, GNN outperforms PHD by a large margin despite using only simple hypothesis management for data association and heuristic rules for track maintenance, since PHD does not use the proper approach for data association, which may not be suitable for the situation in the real-world LiDAR-based 3D MOT. 

In principle, the JPDA and CPHD filters may boost performance 
compared to GNN and PHD, respectively. However, the marginal association probability computation in JPDA \cite{Reid1978AnAF} is NP-hard, and the computational complexity of the cardinality distribution in CPHD increases exponentially with the number of objects. These computation requirements cannot be reached due to the enormous number of detection inputs offered by nuScenes dataset. Due to memory limits, both filters would be automatically terminated during the tracking process.

\subsubsection{Performance of TO-PMB}
\begin{figure}[th]
\centering
        \includegraphics[scale=0.78]{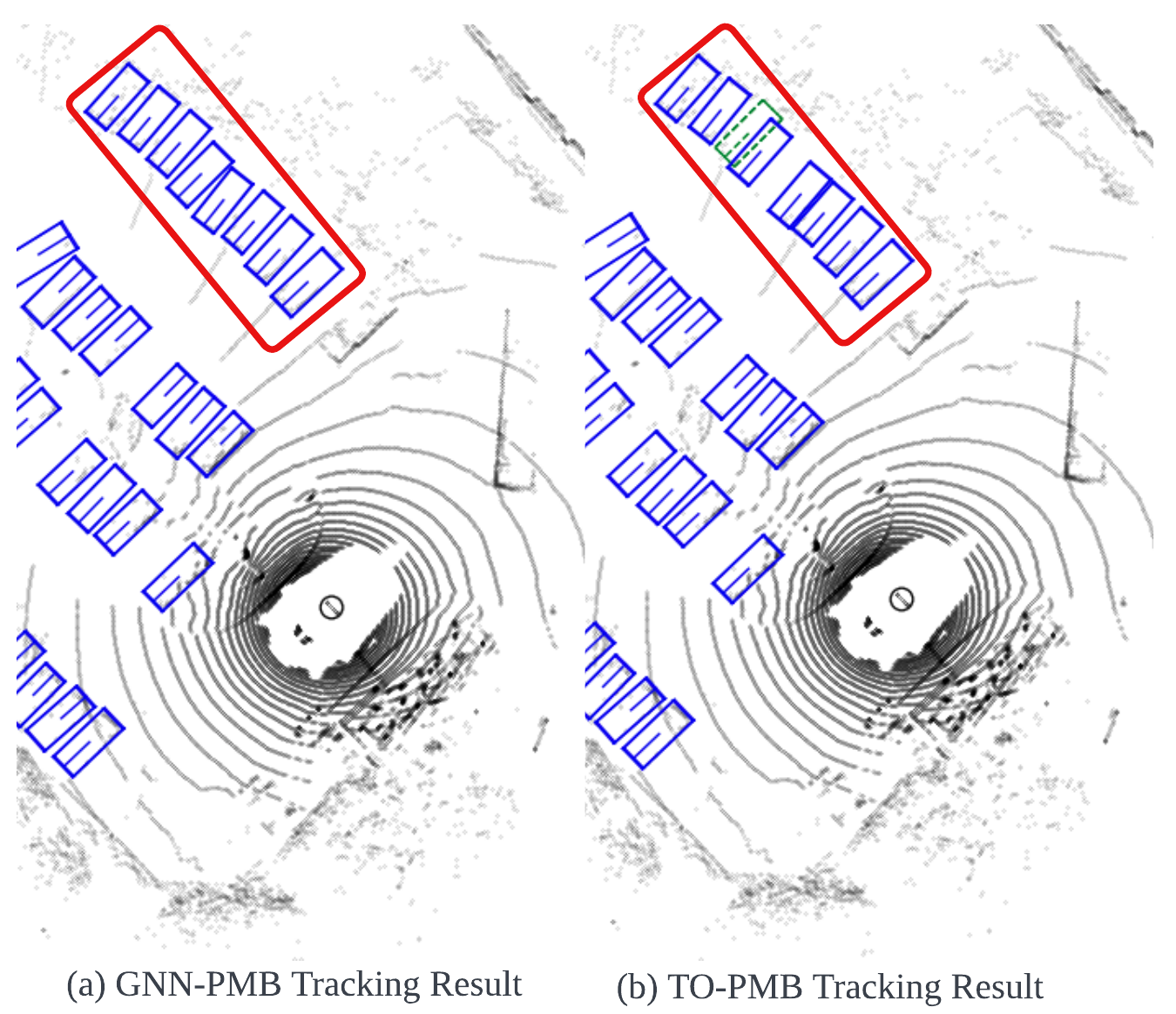}
        \caption{The coalescence of the TP-PMB filter. This figure demonstrates the coalescence problem of the TO-PMB filter by comparing the tracking results of the same time step from both the GNN-PMB tracker and the TO-PMB tracker, where the grey dots are the LiDAR point cloud, the blue bounding boxes are the TP object, and the green bounding box with dashed line is the FN object, i.e., the object indeed exists but has not been tracked. 
        In the highlighted red rectangle area, GNN-PMB tracks all the objects correctly, but TO-PMB incorrectly merges two object estimates and as a result misses one object.
}        \label{TO-PMB}
\end{figure}

TABLE \ref{comparison_between_tracker_frameworks_in_nuScenes_validation_dataset} also shows that the performance of TO-PMB in our experiments is insufficient. One of the reasons is that the TO-PMB suffers from the coalescence problem, which refers to the phenomenon that multiple tracked objects are merged into one tracked object when multiple objects move in close proximity. This effect is demonstrated by Fig. \ref{TO-PMB}, which is taken from a scene where the vehicle goes through a parking lot with tightly parked vehicles. In this particular frame, there should have $8$ parked cars in the studied area, but the TO-PMB filter only tracked seven parked cars as it merged two cars into one.

\begin{figure}[th]
    \centering
    \includegraphics[scale=0.5]{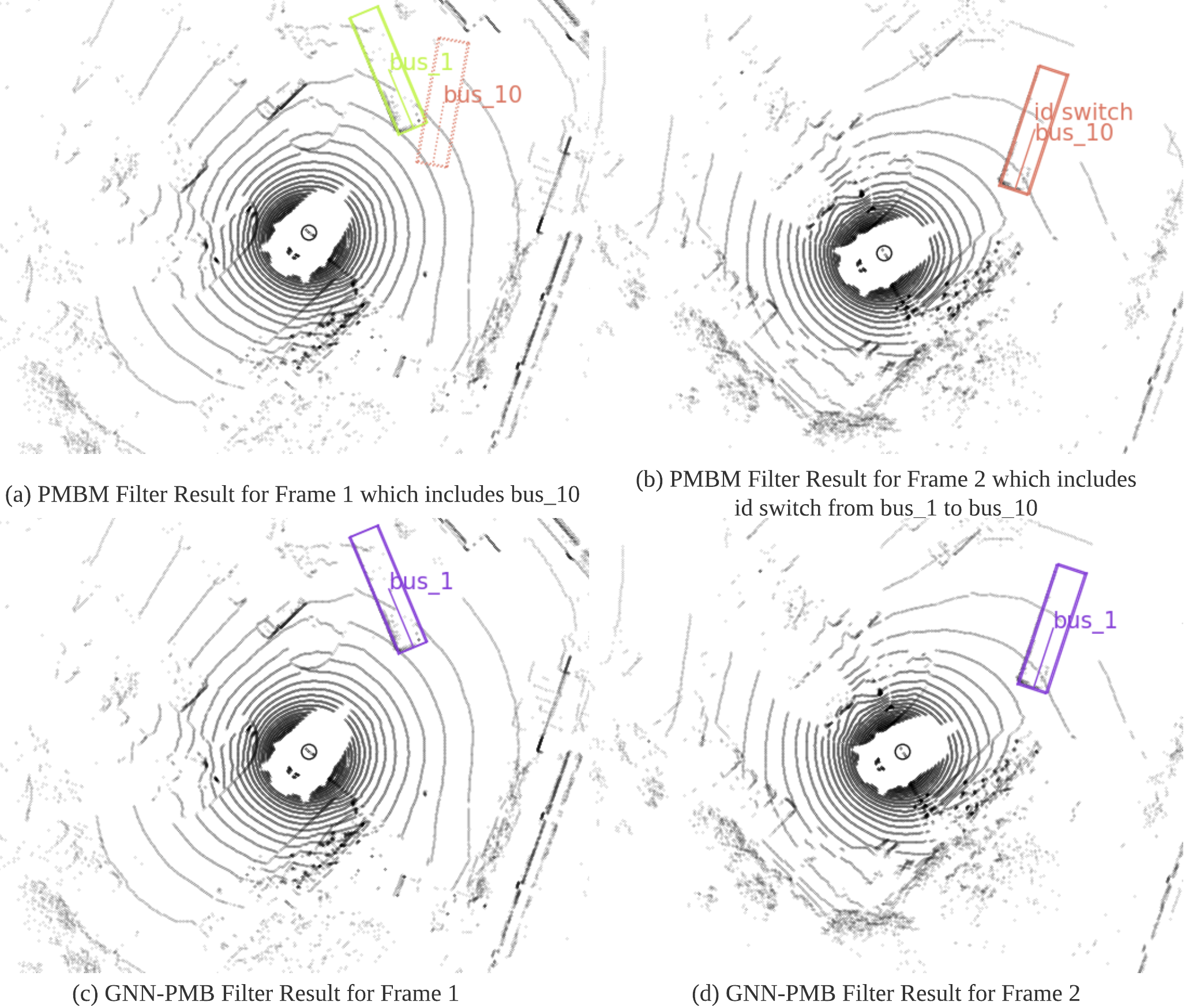}
    \caption{Cascading failure of the PMBM filter. This figure compares the tracking results between the PMBM filter and the GNN-PMB filter. In (a), bus$_{10}$ in dash bounding box is a false alarm track. The track of bus$_{10}$ is sustained by the PMBM filter, leading to an ID switch where the ID of the valid track changes from bus$_{1}$ to bus$_{10}$ in the later time step illustrated by (b). These two errors are a series of cascading failures because the PMBM filter propagates $K$ best global hypotheses. In the GNN-PMB filter, however, only the valid track for bus$_{1}$ is registered, and it correctly maintains the same track ID for bus$_{1}$ in the following time step, as illustrated by (c) and (d).}
    \label{pmbmerror}
\end{figure}

\subsubsection{Performance of PMBM}

The PMBM filter achieves the second best performance in TABLE \ref{comparison_between_tracker_frameworks_in_nuScenes_validation_dataset}, outperforming all other filters but with the AMOTA score $8.769\%$ lower than that of the GNN-PMB with the CenterPoint detector. The performance disparity can be attributed to the fact that incorrect local hypotheses would persist for the PMBM filter. In contrast, only the most likely global hypothesis is propagated in the GNN-PMB filter, thereby decreasig the probability of incorrect local hypotheses being persist. Fig. \ref{pmbmerror} compares the tracking result between the PMBM filter and the proposed GNN-PMB filter. The ground truth contains only one track, bus$_{1}$. However, the PMBM filter erroneously predicted the existence of another track bus$_{10}$. The PMBM filter made a mistake in the first frame, and the error actually persisted to the next frames and resulted in an ID switch. 
%


\subsection{Ablation Study of Parameters for the GNN-PMB Tracker}
\label{ablation}

Since the performance of the proposed GNN-PMB tracker can be tuned by parameters described in Section \ref{paradef}, a set of carefully selected parameters can let the GNN-PMB achieves a reasonably good performance. In this section, we present the parameter tuning process of the proposed GNN-PMB tracker, for the bus class type as an example. According to our experiments, some of the settings have minor influence on the performance of the GNN-PMB tracker 
hence their values are supplied directly 
as follows:
\begin{itemize}
\item{The gating threshold is $\sqrt{40}$ in 2D Mahalanobis distance.}
\item{The clutter rate is set to $0.001$ over the area of FoV.}
\item{The survival probability is set to $0.7$.}
\item{The pruning threshold of existence probability is $10^{-6}$.}
\end{itemize}
Changing the values of the rest parameters, on the other hand, causes significant variations in performance, as shown in TABLE \ref{parameters_ablation}. We elaborate on the details below:

\subsubsection{Detection Score Threshold}

The first row of TABLE \ref{parameters_ablation} indicates that discarding the bounding box with a predetermined detection score threshold would result in a lower AMOTA score. When the detection score threshold is set to be $0.1$, $0.2$, $0.3$, the AMOTA score are $0.707$, $0.672$ and $0.625$ respectively. It should also be noted that the recall decreases, and the number of fragmented trajectories increase as the detection score threshold increases.

The TP score of the three detection thresholds are $83088$, $80971$, and $80601$, respectively. It means that there are $2117$ valid detection with detection score between $0.1$ and $0.2$, and that there are $370$ valid detection with detection score between $0.2$ and $0.3$. As the change of detection score threshold, the number of fragmented trajectories varies as well. Specifically, the number of fragmented trajectories increased from $328$ to $500$, when the detection score threshold increased from $0.2$ to $0.3$. 
On the contrary, the number of fragmented trajectories decreased from $341$ to $328$ when the detection score threshold increases from $0.1$ to $0.2$. When the detection score threshold is set to be $0.2$, there are only $4391$ mostly tracked trajectories, showing a decrease of 215 as compared to $4606$ mostly tracked trajectories when the threshold is set to be $0.1$. Therefore, the decrease in the number of fragmented trajectory is an artifact that fewer mostly tracked trajectories exist.

Our experiment results suggest that crucial information for trajectory continuity is contained in detection with a detection score lower than $0.3$. Trajectory continuity is essential to later modules such as trajectory prediction and motion planning. Our observation shows that rather than applying the law of parsimony and removing the detection with a low detection score, the output of the LiDAR 3D object detector should be provided to the MOT tracker in its entirety. 

\subsubsection{NMS Threshold}

We set the NMS threshold to $0.1$, which allows us to discard the majority of overlapping detected bounding boxes even if only a small overlap occurs, yielding an AMOTA score of $0.707$ with the CenterPoint object detector. In contrast, by setting the NMS threshold to 0.98, almost all the overlapped detected bounding boxes are kept as input to the GNN-PMB tracker, resulting a lower AMOTA score of 0.68. Our observation that a basic NMS preprocess operation can improve the AMOTA score by $0.27$ denotes that only the detected bounding boxes with the least overlap with each other should be used as input to the tracker.

\subsubsection{Poisson Birth Density}

As discussed in section \ref{paradef}, proper initialization of Poisson birth density is critical before tracking the newborn object. When the initial covariance of the Gaussian component increases from $15$ to $100$, the AMOTA score is reduced from $0.699$ to $0.625$, and the related AMOTP score increases from $0.566$ to $0.761$. This is reasonable since, as the covariance grows, the estimation uncertainty would increase, leading to performance degradation.

The weight of Gaussian components in the Poisson intensity also plays a crucial role in tracking newborn objects. When the weights increase, measurement becomes more likely to be construed as originating from a new object than an existing one since the cost of being associated with a new object would decrease. On one extreme, when the weight is too large, every measurement would be interpreted as the initial detection of a new track. On the other extreme, where the weight is too small, the measurements would always be associated with existing tracks, and PPP would initiate no new tracks. According to the ablation study, the AMOTA decreases as the weight decreases, while the number of ID switches varies at a different pace as the weight varies. Consequently, such observation demonstrates that finding a suitable value for the weight of the Gaussian component to fit the statistics of object birth in the nuScenes dataset is essential.

\subsubsection{State Extraction Threshold}
Since the probability of the Bernoulli component signifies how probable an object exists, precisely selecting a reasonable extraction threshold becomes important for the final estimation of object states. According to our ablation investigation, raising the extraction threshold increases AMOTA, MT, and TP while lowering ML, FP, and FN. This result reveals that preserving the Bernoulli components with a high enough existence probability would result in a more accurate track estimation.

\begin{table*}[ht]
\caption{Tracking results of proposed method and different model-based trackers using LiDAR on nuScenes} 
\label{comparison_with_sota_model_based_trackers}
\begin{center}
\begin{threeparttable}
\begin{tabular}{ l|ccccccccc}
  \hline
  Method & AMOTA$\uparrow$ & AMOTP$\downarrow$ & MT$\uparrow$ & ML$\downarrow$ & TP$\uparrow$ & FP$\downarrow$ & FN$\downarrow$ & IDS$\downarrow$ & FRAG$\downarrow$ \\
  \hline
AB3DMOT (IROS 2020)* \cite{AB3DMOT} & 0.151 & 1.501 & 1006 & 4428 & 34808 & 15088 & 75730 & 9027 & 2557 \\
StanfordIPRL-TRI (NeurIPS Workshop 2019)* \cite{StanfordIPRL-TRI} & 0.550 & 0.798 & 4294 & 2184 & 85399 & 17533 & 33216 & 950 & 776 \\
RFS-M3 (ICRA 2021)* \cite{Pang20213DMT} & 0.619 & 0.752 & 5107 & 1878 & 90872 & 16728 & 27168 & 1525 & 856 \\
CBMOT-LiDAR (IROS 2021)* \cite{CBMOT} & 0.649 & 0.592 & 5319 & 1966 & 94916 & \textbf{16469} & 24092 & 557 & 450 \\
SimpleTrack (Arxiv 2021)* \cite{SimpleTrack} & 0.668 & \textbf{0.550} &5476 & 1780 & 95539 & 17514 & 23451 & 575 & 591 \\
BPTracker (Proceedings of the
IEEE 2018)* \cite{BPTracker} & 0.646 & 0.606 & 5186 & 2259 & 95053 &
18581 & 24358 & \textbf{154} & \textbf{221} \\
ImmortalTracker (Arxiv 2021)* \cite{ImmortalTracker} & 0.677 & 0.599 & 5565 & 1669 & 96584 & 18012 & 21661 & 320 & 477 \\
GNN-PMB (Our)* & \textbf{0.678} & 0.560 & \textbf{5698} & \textbf{1622} & \textbf{97274} & 17071 & \textbf{21521} & 770 & 431 \\
\hline
PF-MOT tracker (ICRA 2022)** \cite{PF-MOT} & 0.682 & N/A & N/A & N/A & N/A & N/A & N/A & N/A & N/A \\
GNN-PMB (Ours)** & \textbf{0.707} & \textbf{0.560} & \textbf{4608} & \textbf{1347} & \textbf{83134} & \textbf{12362} & \textbf{18113} & \textbf{650} & \textbf{345} \\
\hline
\end{tabular}
\begin{tablenotes}
        \footnotesize
        \item[*] 
        The metrics are reported on the nuScenes test set.
        \item[**] The metrics are reported on the nuScenes validation set.
\end{tablenotes}
\end{threeparttable}
\end{center}
\end{table*}

\begin{table*}[ht]
\caption{Tracking results of proposed method and different learning-based trackers using LiDAR on nuScenes} 
\label{comparison_with_learning_based_trackers}
\begin{center}
\begin{threeparttable}
\begin{tabular}{ l|ccccccccc}
  \hline
  Method & AMOTA$\uparrow$ & AMOTP$\downarrow$ & MT$\uparrow$ & ML$\downarrow$ & TP$\uparrow$ & FP$\downarrow$ & FN$\downarrow$ & IDS$\downarrow$ & FRAG$\downarrow$ \\
\hline
SimTrack (ICCV 2021)* \cite{SimTrack}  & 0.645 & 0.681 & 5063 & 1986 & 92093 & 17443 & 26430 & 1042 & 472 \\
OGR3MOT (IEEE RAL 2022)* \cite{OGR3MOT}  & 0.656 & 0.620 & 5278 & 2094 & 95264 & 17877 & 24013 & 288 & 371 \\
NEBP (Arxiv 2022)* \cite{NEBP}  & 0.673 & 0.586& 5380 & 2126 & 97023 & 19535 & 22380 & \textbf{162} & \textbf{256} \\
GNN-PMB (Our)* & \textbf{0.678} & \textbf{0.560} & \textbf{5698} & \textbf{1622} & \textbf{97274} & \textbf{17071} & \textbf{21521} & 770 & 431 \\
\hline
TransMOT (IEEE IV 2022)** \cite{TransMOT} & 0.674 & 0.754 & 2096 & N/A & N/A & 9449 & 14071 & 1403 & N/A \\
GNN-PMB (Ours)** & \textbf{0.849} & \textbf{0.387} & \textbf{2762} & \textbf{668} & \textbf{49182} & \textbf{6140} & \textbf{8791} & \textbf{344} & \textbf{170} \\
\hline
\end{tabular}
\begin{tablenotes}
        \footnotesize
        \item[*] The metrics are reported on the nuScenes test set.
        \item[**] The metrics are reported on the nuScenes validation set for car.
\end{tablenotes}
\end{threeparttable}
\end{center}
\end{table*}

\begin{table*}[tht]
\caption{Tracking results of proposed method and different trackers using LiDAR and camera fusion on nuScenes test set} 
\label{comparison_with_LiDAR_camera_fusion_trackers}
\begin{center}
\begin{threeparttable}
\begin{tabular}{ l|ccccccccc}
  \hline
  Method & AMOTA$\uparrow$ & AMOTP$\downarrow$ & MT$\uparrow$ & ML$\downarrow$ & TP$\uparrow$ & FP$\downarrow$ & FN$\downarrow$ & IDS$\downarrow$ & FRAG$\downarrow$ \\
\hline
Probabilistic3DMM (ICRA 2021)* \cite{Probabilistic3DMM} & 0.655 & 0.617 & 5494 & \textbf{1557} & 95199 & 18061 & 23323 & 1043 & 717 \\
EagerMOT (ICRA 2021)* \cite{EagerMOT} & 0.677 & 0.550 & 5303 & 1842 & 93484 & 17705 & 24925 & 1156 & 601 \\
CBMOT (IROS 2021)* \cite{CBMOT} & 0.676 & \textbf{0.518} & 5420 & 1654 & 96028 & 21604 & 22828 & \textbf{709} & 1015 \\
AlphaTrack (IROS 2021)* \cite{AlphaTrack} & \textbf{0.693} & 0.585 & 5560 & 1744 & 95851 & 18421 & 22996 & 718 & 480 \\
GNN-PMB (Ours)** & 0.678 & 0.560 & \textbf{5698} & 1622 & \textbf{97274} & \textbf{17071} & \textbf{21521} & 770 & \textbf{431} \\
\hline
\end{tabular}
\begin{tablenotes}
        \footnotesize
        \item[*] All trackers are based fusion of LiDAR and camera.
        \item[**] Our tracker is based on LiDAR only.
\end{tablenotes}
\end{threeparttable}
\end{center}
\end{table*}

\subsection{Comparison with Other State-Of-The-Art Methods}
\label{sotacomp}

\subsubsection{Performance Comparison with State-of-The-Art Model-based LiDAR Trackers using LiDAR Only}
\label{modelcomp}

In the nuScenes test dataset, our proposed LiDAR only tracker, GNN-PMB, is compared against various model-based LiDAR only trackers, and the results are reported in TABLE \ref{comparison_with_sota_model_based_trackers}. Among the model-based trackers, the proposed GNN-PMB tracker receives the highest AMOTA score of 0.678. In terms of AMOTP, the GNN-PMB tracker achieves a score of 0.560, which is only 0.01 lower than that of SimpleTrack, and it is the second-best AMOTP score amongst all compared model-based trackers. However, it should be noted that all the other trackers use all the 3D information provided by the object detector as input of the tracker, whereas the GNN-PMB tracker only utilizes the x and y coordinates in the global frame as input.

In addition, our proposed method also outperforms all the model-based trackers in MT, ML, TP, and FN, with the FP score coming in second only to CBMOT-LiDAR, and notably, achieving an MT score of 5698. Even with lower frequency input (at 2Hz), the GNN-PMB tracker still managed to track 4.05\% more tracks than SimpleTrack, which employed detection at 10Hz as the input. Moreover, the GNN-PMB tracker is even superior to ImmortalTracker, which refines the trajectory with more accumulated information in the future through post-processing, though it is impractical in the real-world system for online tracking. Another recently proposed LiDAR-only model-based tracker, the PF-MOT tracker, only reported performances partially in the nuScenes validation dataset excluding the test dataset, limiting the comparison on the validation dataset. The result shows that the GNN-PMB tracker achieves an AMOTA of 0.707, which is higher than the 0.682 AMOTA achieved by the PF-MOT tracker.

The drawbacks of the GNN-PMB tracker are the higher IDS and FRAG compared to BPTracker, which has only 154 ID switches and 221 fragments. However, considering relatively low values in MT, the low ID switches achieved by the BPTracker could be partially due to an artifact from having less mostly tracked tracks.

\subsubsection{Performance Comparison with State-Of-The-Art Data-driven Trackers using LiDAR Only}
\label{learning}
We also compare our proposed GNN-PMB with data-driven state-of-the-art trackers. As shown by TABLE \ref{comparison_with_learning_based_trackers}, our proposed GNN-PMB tracker achieves the best performance in all evaluation metrics other than IDS and FRAG, when it is compared with the data-driven trackers. In particular, our proposed GNN-PMB tracker achieves an AMOTP score of 0.56, which is 0.026 better than the NEBP tracker, the second-best tracker among all the learning-based trackers. Furthermore, the AMOTP score indicating that the GNN-PMB tracker can provide more accurate position information than the NEBP tracker.

\subsubsection{Performance Comparison with State-Of-The-Art Trackers using LiDAR and Camera Fusion}
\label{camera}

To further demonstrate the advantages of our proposed GNN-PMB tracker, we present the comparison between the proposed GNN-PMB and the state-of-the-art trackers using LiDAR and camera fusion. It is worth pointing out that our proposed GNN-PMB tracker using LiDAR only, still can obtain comparable tracking performance on AMOTA and AMOTP and it is even superior on MT, TP, FP, and FN, as shown in TABLE \ref{comparison_with_LiDAR_camera_fusion_trackers}. Such results show an enormous potential to increase performance even further when the GNN-PMB tracking framework is extended into settings with the fusion of LiDAR and camera.

\section{Conclusion and Future Work}
\label{conclusion}

A systematic comparison is provided among different random vector-based Bayesian and RFS filters on the nuScenes dataset. 
Based on the analysis, a simple but effective online multi-object tracker GNN-PMB is proposed.
Due to the simple structure of the RFS framework, the proposed GNN-PMB tracker requires no additional module, heuristic trick, or data association and track maintenance rule. 
Thus, it is simple to tune, and it achieves the state-of-the-art performance on nuScenes dataset. Its performance can be further improved with simple modifications, such as providing 3D state information as input, using both LiDAR and camera as sensor modalities, and including a meta-learning-based parameter auto-tuning module, etc.
In our future work, downstream tasks such as trajectory prediction and motion planning will be incorporated to evaluate the multi-object trackers, such that a more robust multi-object tracker in an end-to-end manner can be designed.



\small


\vspace{-10mm}
\begin{IEEEbiography}[{\includegraphics[width=1in,height=1.25in,clip,keepaspectratio]{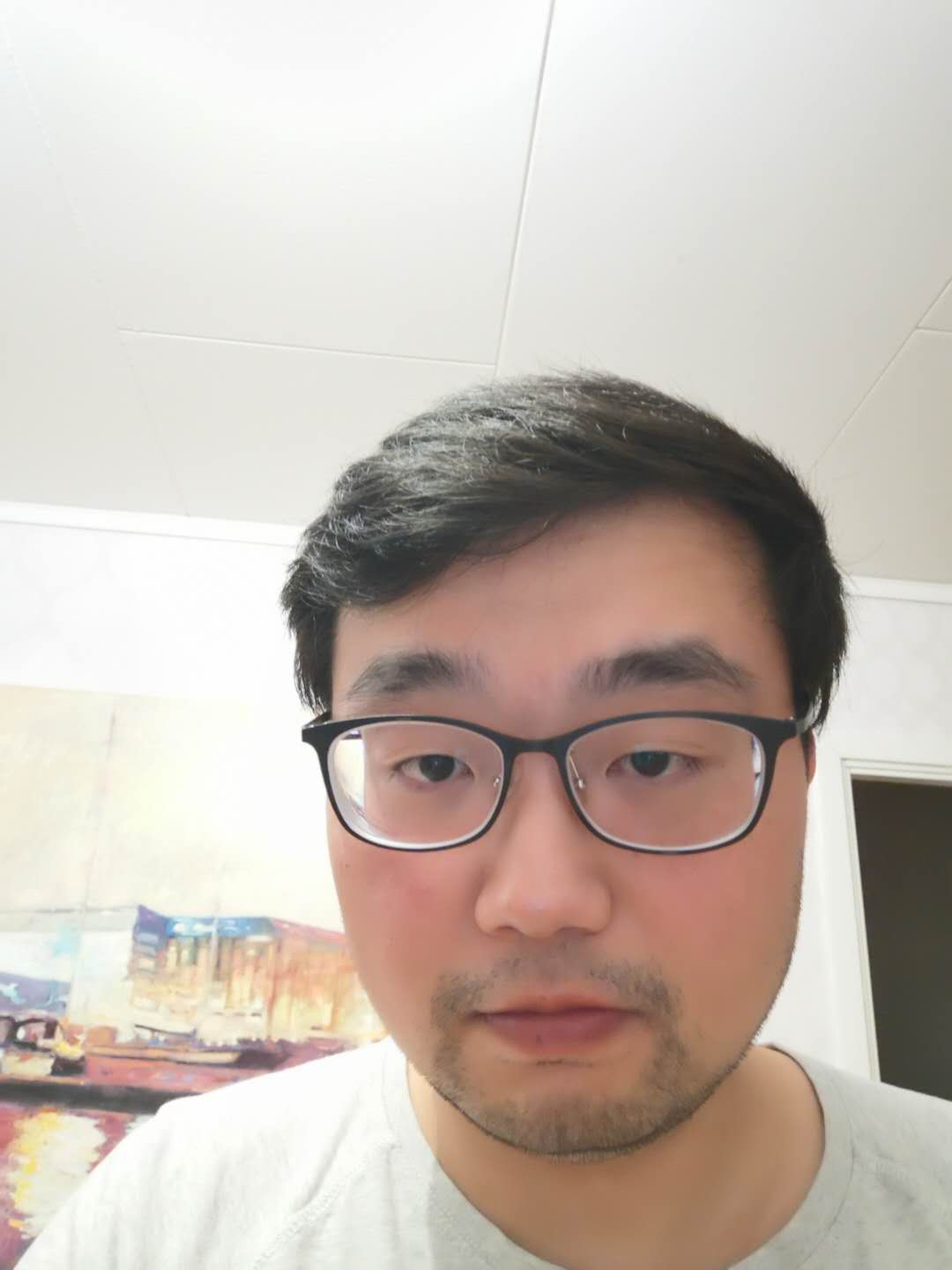}}]{Jianan Liu} received his B.Eng. degree in Electronics and Information Engineering from Huazhong University of Science and Technology, Wuhan, China, in 2007. He received his M.Eng. degree in Telecommunication Engineering from the University of Melbourne, Australia, and his M.Sc. degree in Communication Systems from Lund University, Sweden, in 2009 and 2012, respectively.
Jianan has over ten years of experience in software and algorithm design and development. He has held senior R\&D roles in the AI consulting, automotive, and telecommunication industries.
His research interests include applying statistical signal processing and deep learning for medical image processing, wireless communications, IoT networks, indoor sensing, and outdoor perception using a variety of sensor modalities like radar, camera, LiDAR, WiFi, etc.
\end{IEEEbiography}

\vspace{-10mm}
\begin{IEEEbiography}[{\includegraphics[width=1in,height=1.25in,clip,keepaspectratio]{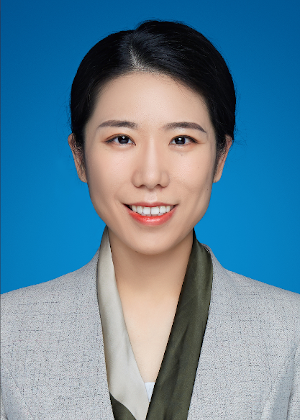}}]{Liping Bai} received her M.Eng. degree in Instrument Engineering from Nanjing University of Posts and Telecommunications, Nanjing, China, in 2021. She is currently pursuing her Ph.D. in Automation and Control Engineering at Beihang University, Beijing, China. Her research interest is the intersection between control theory and machine learning.
\end{IEEEbiography}

\vspace{-10mm}
\begin{IEEEbiography}[{\includegraphics[width=1in,height=1.25in,clip,keepaspectratio]{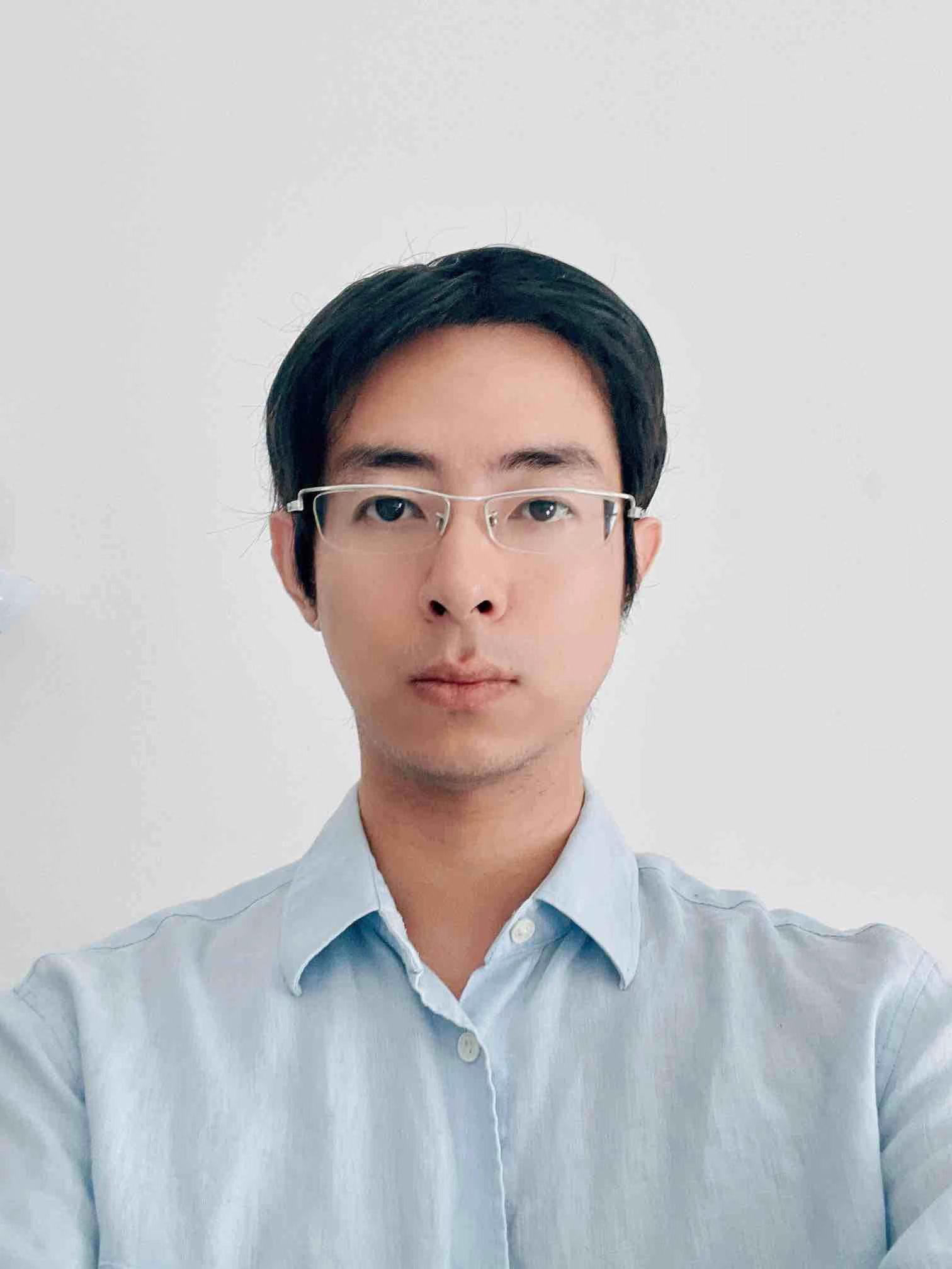}}]{Yuxuan Xia} received the M.Sc. degree in communication engineering and the Ph.D. degree in signal processing from Chalmers University of Technology, Gothenburg, Sweden, in 2017 and 2022, respectively. He is currently a Postdoctoral researcher with the Department of Electrical Engineering, Chalmers University of Technology. His main research interests include multi-object tracking and sensor fusion, especially for extended objects. He has co-organized tutorials on multi-object tracking at the Fusion 2020, 2021 and 2022 conferences.
\end{IEEEbiography}

\vspace{-10mm}
\begin{IEEEbiography}[{\includegraphics[width=1in,height=1.25in,clip,keepaspectratio]{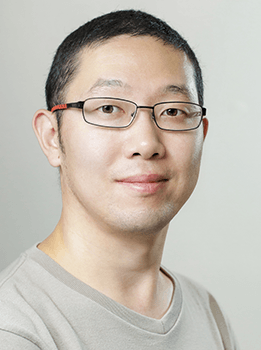}}]{Tao Huang}
(Senior Member, IEEE) received his Ph.D. degree in Electrical Engineering from The University of New South Wales, Sydney, Australia. 
He received his M.Eng. degree in Sensor System Signal Processing from The University of Adelaide, Adelaide, Australia. He received his B.Eng. Degree in Electronics and Information Engineering from Huazhong University of Science and Technology, Wuhan, China. 
Currently, Dr Huang is a lecturer in Electronic Systems and IoT Engineering at James Cook University, Cairns, Australia. 
He was an Endeavour Australia Cheung Kong Research Fellow, a visiting scholar at The Chinese University of Hong Kong, a research associate at the University of New South Wales, and a postdoctoral research fellow at James Cook University. 
He has co-authored a Best Paper Award from the 2011 IEEE WCNC, Cancun, Mexico.
He is a co-inventor of one patent on MIMO systems.  
He has served in several international conferences as TPC chair, track chair, program vice chair, and local chair. 
He is currently serving as the MTT-S/Com Vice-Chair and Young Professionals Affinity Group Chair for the IEEE Northern Australia Section. 
Before academia, Dr Huang held various positions in the industry, such as senior software engineer, senior data scientist, project team lead, and technical authority. 
His research interests include deep learning, smart sensing, computer vision, pattern recognition, wireless communications, and IoT security.
\end{IEEEbiography}

\vspace{-10mm}
\begin{IEEEbiography}[{\includegraphics[width=1in,height=1.25in,clip,keepaspectratio]{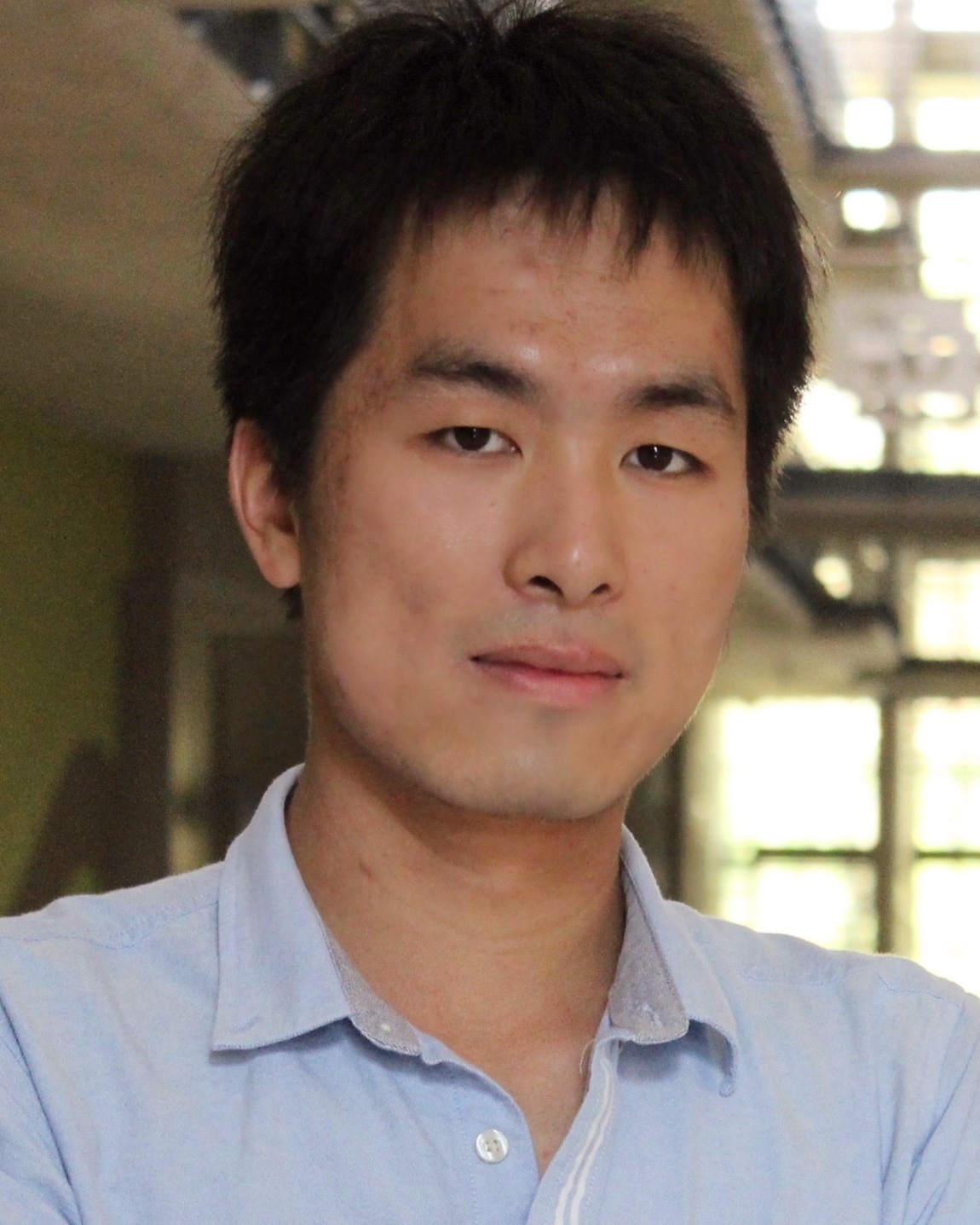}}]{Bing Zhu}(Member, IEEE) was born in Wuhan, P.R. China. He received his B.S and Ph.D. degrees in Control Theory and Applications from Beihang University, Beijing, PR China, in 2007 and 2013, respectively. He was with University of Pretoria, Pretoria, South Africa, as a postdoctoral fellow supported by Vice-Chancellor Postdoctoral Fellowship from 2013 to 2015. He was with Nanyang Technological University, Singapore, as a research fellow from 2015 to 2016. He joined Beihang University, Beijing, PR China as an associate professor in 2016.

Dr. Zhu serves as an Associate Editor for Acta Automatica Sinica. His research interests include model predictive control, smart sensing for UAV and UGV, and demand-side management for new energy systems.
\end{IEEEbiography}

\vspace{-10mm}
\begin{IEEEbiography}[{\includegraphics*[width=1in,height=1.25in,clip,keepaspectratio]{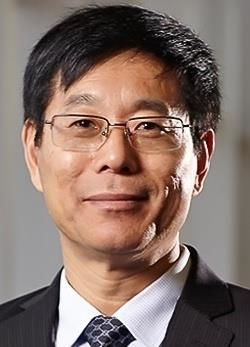}}]{Qing-Long Han} (Fellow, IEEE) received the B.Sc. degree in Mathematics 
from Shandong Normal University, Jinan, China, in 1983, and the M.Sc. and Ph.D. degrees in Control Engineering from East China University of Science and Technology, 
Shanghai, China, in 1992 and 1997, respectively.

Professor Han is Pro Vice-Chancellor (Research Quality) and a Distinguished Professor at Swinburne University of Technology, Melbourne, Australia. He held various academic 
and management positions at Griffith University and Central Queensland University, Australia. His research interests include networked control systems, multi-agent systems, 
time-delay systems, smart grids, unmanned surface vehicles, and neural networks.

Professor Han was awarded The 2021 Norbert Wiener Award (the Highest Award in systems science and engineering, and cybernetics) and The 2021 M. A. Sargent Medal 
(the Highest Award of the Electrical College Board of Engineers Australia). He was the recipient of The 2021 IEEE/CAA Journal of Automatica Sinica Norbert Wiener Review Award, 
The 2020 IEEE Systems, Man, and Cybernetics (SMC) Society Andrew P. Sage Best Transactions Paper Award, The 2020 IEEE Transactions on Industrial Informatics Outstanding Paper Award, 
and The 2019 IEEE SMC Society Andrew P. Sage Best Transactions Paper Award.

Professor Han is a Member of the Academia Europaea (The Academy of Europe). He is a Fellow of The International Federation of Automatic Control (IFAC) and a Fellow of The Institution of Engineers Australia (IEAust).  
He is a Highly Cited Researcher  in both Engineering and Computer Science (Clarivate Analytics). He has served as an AdCom Member of IEEE Industrial Electronics Society (IES), a Member of IEEE IES Fellows Committee, 
and Chair of IEEE IES Technical Committee on Networked Control Systems. He is Co-Editor-in-Chief of IEEE Transactions on Industrial Informatics, Deputy Editor-in-Chief of IEEE/CAA JOURNAL OF AUTOMATICA SINICA, Co-Editor of Australian Journal of Electrical and Electronic Engineering, 
an Associate Editor for 12 international journals, including the IEEE TRANSACTIONS ON CYBERNETICS, IEEE INDUSTRIAL ELECTRONICS MAGAZINE, Control Engineering Practice, and Information Sciences, and a Guest Editor for 14 Special Issues.
\end{IEEEbiography}

\clearpage
\appendix
\subsection{Hypothesis Management in GNN-PMB, PMB, and PMBM}

We show that the hypothesis management scheme in proposed GNN-PMB is what differentiates our method from the well-known PMBM filter. PMBM filter propagates the $N_h$ best global hypotheses from the previous frame to the current frame. GNN-PMB filter propagates the best global hypothesis from the previous frame to the current frame. In terms of implementation, to best illustrate the difference between the PMBM filter and the GNN-PMB filter, we put the pseudo code of the PMBM filter with that of the proposed GNN-PMB filter side by side. The procedure of PMBM presented in \cite{GarcaFernndez2018PMBM} is explained by Algorithm \ref{alg:PMBM}, and the procedure proposed in our method, GNN-PMB, is explained by Algorithm \ref{alg:GNNPMB}.

\begin{algorithm}[ht]
\caption{{\label{alg:PMBM}{Pseudo-code
for the PMBM filter}}}

{\fontsize{9}{9}\selectfont

\textbf{Input:} Parameters of the PMBM posterior at the previous time
step and measurement
set $Z$ at the current time step

\textbf{Output: }Parameters of the PMBM posterior at the current time step

\begin{algorithmic}[ht]     
\State $\phantom{}$ {\Comment{Prediction}}
\For{$p=1$ to $P$} \Comment{We go through all the Poisson components}
\State - Perform prediction step
\EndFor

\For{$b=1$ to $B$} \Comment{We go through all the Bernoulli components}
\State - Perform prediction step
\EndFor

\State $\phantom{}$ {\Comment{Update}}

\For{$z\in Z$} \Comment{Targets detected for first time}

\State - Perform ellipsoidal gating of $z$ w.r.t. Gaussian components
of Poisson prior

\If{$z$ meets ellipsoidal gating for at least one component}

\State - Create a new Bernoulli component

\EndIf

\EndFor

\For{$i=1$ to $n$} \Comment{We go through all possible targets}

\For{$j_{i}=1$ to $l_{i}$} \Comment{$l_{i}$ is the number of local hypotheses for possible target $i$}

\State - Create new misdetection hypothesis

\State - Perform gating on $Z$ and create new detection hypotheses

\EndFor

\EndFor

\For{all $j$} \Comment{We go through all global hypotheses propagated from previous time step}

\State - Create cost matrix

\State - Run Murty's algorithm to select $H_{k}=\left\lceil N_{h}\cdot w_{j}\right\rceil $
new global hypotheses

\EndFor

\State - Estimate target states

\State $\phantom{}$ {\Comment{Pruning}}

\State - Prune the Poisson part by discarding components whose weight
is below a threshold 

\State - Prune global hypotheses by keeping the highest $N_{h}$
global hypotheses

\State - Remove Bernoulli components whose existence probability
is below a threshold or do not appear in the pruned global hypotheses

\end{algorithmic}}
\end{algorithm}

\begin{algorithm}[h!]
\caption{{\label{alg:GNNPMB}{Pseudo-code
for the GNN-PMB filter}}}

{\fontsize{9}{9}\selectfont

\textbf{Input:} Parameters of the GNN-PMB posterior at the previous time
step, and measurement
set $Z$ at the current time step

\textbf{Output: }Parameters of the GNN-PMB posterior at the current time
step

\begin{algorithmic}[h]     

\State $\phantom{}$ {\Comment{Prediction}}
\For{$p=1$ to $P$} \Comment{We go through all the Poisson components}
\State - Perform prediction step
\EndFor

\For{$b=1$ to $B$} \Comment{We go through all the Bernoulli components.}
\State - Perform prediction step
\EndFor

\State $\phantom{}$ {\Comment{Update}}

\For{$z\in Z$} \Comment{Targets detected for first time}

\State - Perform ellipsoidal gating of $z$ w.r.t. Gaussian components
of Poisson prior

\If{$z$ meets ellipsoidal gating for at least one component}

\State - Create a new Bernoulli component

\EndIf

\EndFor
\For{$i=1$ to $n$} \Comment{We go through all possible targets}
\State - Create new misdetection hypothesis.
\State - Perform gating on $Z$ and create new detection hypotheses
\EndFor

\State - Create Cost Matrix for the best global hypothesis propagated from previous time step

\State - Run Hungarian's algorithm to select $H_{k} = 1$
new global hypothesis

\State - Estimate target states

\State $\phantom{}$ {\Comment{Pruning}}

\State - Prune the Poisson part by discarding components whose weight
is below a threshold

\State - Prune global hypotheses by keeping the highest 1
global hypotheses

\State - Remove Bernoulli components whose existence probability
is below a threshold or do not appear in the pruned global hypotheses
\end{algorithmic}}
\end{algorithm}

Specifically, the difference concerning the hypothesis management schemes shown in the two algorithms(pseudo codes as shown above), are specified as the following:

\begin{itemize}
    \item {For the PMBM filter, target $i$ has $l_i$ associated local hypotheses, where $l_i$ is a varying number. In the GNN-PMB filter, on the other hand, each target only has 1 associated local hypothesis, which is part of the only one best global hypothesis propagated from previous time step. Because of this difference, there is no for loop associated with the GNN-PMB filter.}
    
    \item{For the PMBM filter, there are j global hypotheses propagated from previous time step, each with its new cost matrix. Then $H_{k}$ best global hypotheses are selected from the cost matrix. Number $H_{k}$ is computed by multiplying $N_h$, which is the maximum allowed number of global hypotheses at current time step, with $w_j$, which is the weight of the global hypothesis. The GNN-PMB filter on the other hand only has 1 single global hypothesis(which is the best one at previous time step) propagated from previous time step, therefore there is no looping over j global hypotheses. For that single global hypothesis propagated from previous time step as prior, a cost matrix is generated, and the Hungarian algorithm is applied to select the best global hypothesis association scheme for current time step based on the cost matrix.}
\end{itemize}

Notice that for the PMBM filter, if we set the $N_h$ to be 1, then the Murty algorithm would only select the best set of associations. This is equivalent to the Hungarian algorithm. In our implementation, that is exactly what we did, setting the maximum allowed global hypothesis $N_h$ to be 1. Under this implementation, the parts highlighted in the PMBM filter pseudo code are the same as the parts highlighted in the GNN-PMB filter pseudo code, since $l_i$=1 and j=1. 

\begin{figure}[ht]
    \centering
    \includegraphics[scale=0.35]{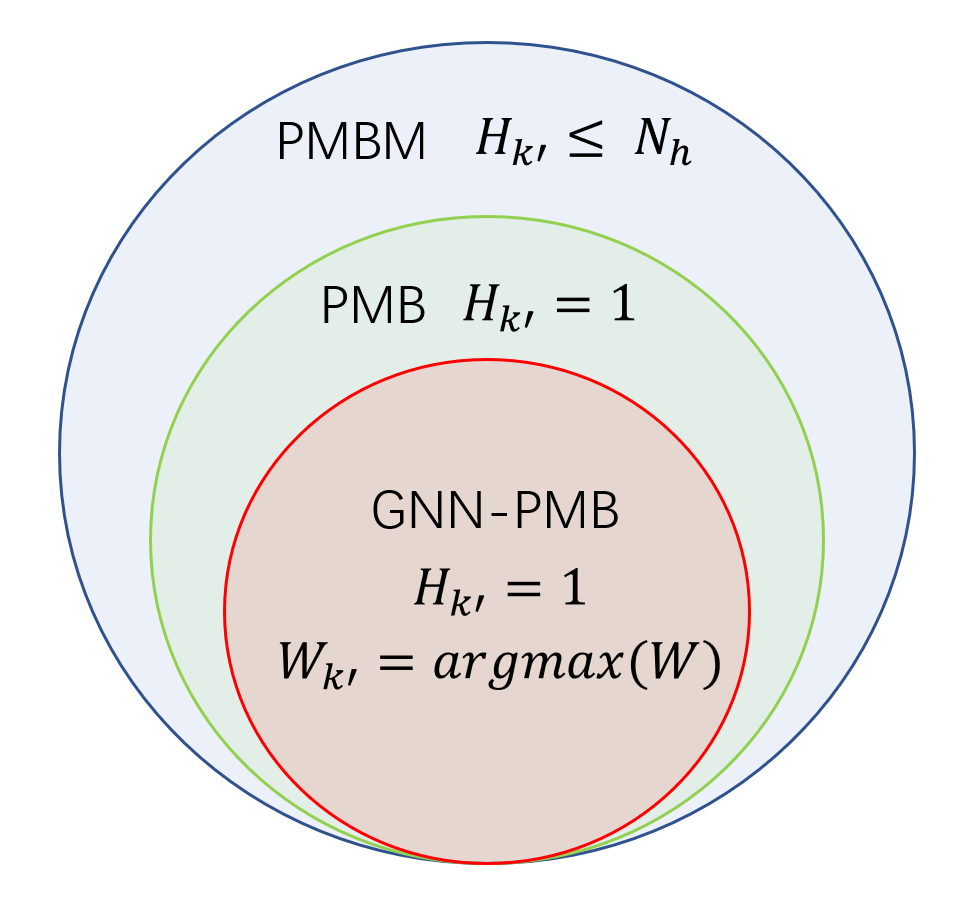}
    \caption{Difference in Hypothesis Management for the PMBM filter, the PMB filter and the GNN-PMB filter.}
    \label{Diff_Hypothesis_Management_PMBM_PMB_GNNPMB}
\end{figure}

The brief explanation of differences between hypothesis management in GNN-PMB, PMB, and PMBM are concluded as below:

For the hypothesis management procedure in the PMBM filter, $H_{k}$ global hypotheses are generated by applying the Murty algorithm to the cost matrices. In equation \eqref{eq_mbm} in Section \ref{background}, $w^h$ indicates the weight of the $h$-th multi-Bernoulli component, and the sum of the weights adds up to one. The maximum allowable number of global hypotheses, $N_h$, is a tunable parameter for the PMMB filter. For instance, when $N_h$ is set to be 100, then $H_{k'}$ has to be a number smaller than or equal to 100. If we only propagate one multi-Bernoulli component, i.e., $H_{k}=1$, then the PMBM filter reduces to a PMB filter \cite{williams2015marginal}. Notice we did not specify how the one multi-Bernoulli component is chosen. However, if we require that the one propagated multi-Bernoulli component must represent the best bipartite matching scheme, then the PMBM filter reduces to the proposed GNN-PMB filter. This relationship between the PMBM filter, the PMB filter and the GNN-PMB filter is demonstrated by Figure \ref{Diff_Hypothesis_Management_PMBM_PMB_GNNPMB}.

In terms of implementation, in the PMBM filter, if we specify that the Murty algorithms only generate the best global hypothesis, then it is equivalent to the Hungarian algorithm. This can be easily accomplished by specifying that $N_h=1$. 

\subsection{Detailed Explanation for the Evaluation Metrics in nuScenes}
\label{metrics}

\begin{figure}[ht]
\begin{center}
\includegraphics[scale=0.70]{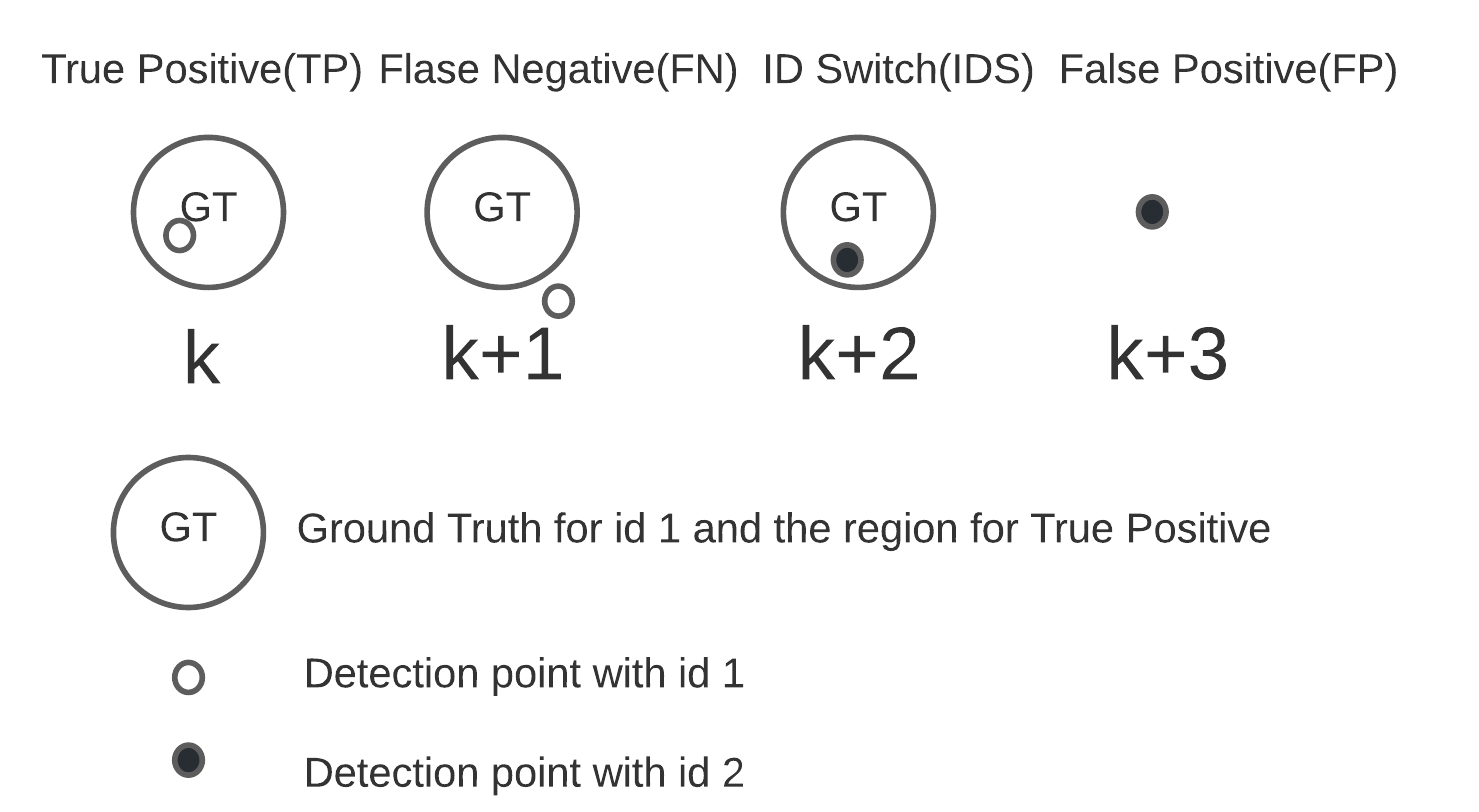}
\caption{Tracking evaluation metrics. This figure explains the definition of TP, FN, IDS and FP. True positive (TP) is the number of detection that fall within the valid region of its corresponding ground truth. False negative (FN) is the number of detection that fall outside the valid region of its corresponding ground truth. ID switch (IDS) is the number of detection that fall within the valid region of its corresponding ground truth, but the IDs assigned to the detection differ from that of the previous frame. False positive (FP) is the number of detection not associated with any ground truth. For the nuScenes dataset, the valid region is defined as a region around the ground truth position with a maximum Euclidean distance of $3$. }
\label{fig:ev}
\end{center}
\end{figure}

Multi-object tracking accuracy(MOTA) and multi-object tracking precision(MOTP) is the most widely used metrics for MOT evaluation. To define MOTA and MOTP, some key secondary metrics are introduced in Fig. \ref{fig:ev}. Mostly tracked (MT) Tracks are tracks with more than $80\%$ TP detection. Mostly lost (ML) Tracks are tracks with less than $20\%$ TP detection. The detailed definitions of secondary metrics can be found on the organiser's website: {\url{https://www.nuScenes.org/tracking}}.

MOTA of a given frame is defined by the equation 
\begin{align}\text{MOTA} = 1-\frac{\sum{\text{FN}+\text{FP}+\text{IDS}}}{\text{GT}}, \label{mota}\end{align}
where $\text{GT}$ denotes the number of ground truth. 

MOTP of a given frame is defined by
\begin{equation}
    \text{MOTP} = \frac{\sum_i{d_i}}{\text{TP}},
\end{equation}
where $d_i$ is the 3D Euclidean distance between the $i^{th}$ TP detection position and its corresponding ground truth position.

However, the nuScenes tracking challenge does not use MOTA and MOTP as the primary evaluation metrics for the tracking performance. The challenge differentiates tracking results based on the average multi-object tracking accuracy (AMOTA).

To define AMOTA, the tracking score and the recall must be defined first. The object detector assigns a detection score between 0 and 1 to each bounding box. The score indicates the likelihood of the identified classification. For instance, a 3D bounding box might be assigned a detection score of 0.3 for it to be a bounding box for a car and a score of 0.1 for it to be a bounding box for a truck. A tracking score is an object prediction score between 0 and 1 for the class identified. For the trackers evaluated in this paper, the tracking score is the same as the detection score.

The tracking score threshold for the AMOTA calculation is dynamically computed for each tracking result. For instance, for a given tracking result, the first recall greater than $0.1$ is $0.104$. It appears when the tracking score threshold is set to be $0.75$. The highest recall is $0.85$. It appears when the tracking score threshold is set to be $0.11$. The $n$ evaluated recalls are evenly spaced between $0.104$ and $0.85$, and the corresponding tracking score threshold would be used to evaluate secondary metrics under that recall. Of the evenly spaced recalls between $0.104$ and $0.85$, the recall with the highest corresponding MOTA would be used to generate all the secondary metrics of this tracking result.

Recall of a given tracking score threshold is defined by
\begin{equation}
    \text{r(ts)} = \frac{\text{TP}_{\text{ts}}}{\text{GT}},    
\end{equation}
where ${ts}$ is the tracking score threshold. $ TP_{ts}$ is the number of True Positives when only detection with a tracking score higher than a specified tracking score threshold is considered valid.

The objective of AMOTA is to evaluate the tracking performances under different recalls. For a given frame, AMOTA is defined by
\begin{equation}
    \text{AMOTA} = \frac{1}{\text{n-1}} \sum_{\text{r} \in \{ \frac{1}{\text{n-1}},  \frac{2}{\text{n-1}}, ... , 1\} } \text{MOTAR}  
\end{equation}
where $n$ is the number of evaluated recalls, and $r$ is the recall. For the nuScenes dataset, $n$ is specified to be 40.

The MOTAR is the MOTA under a given recall. It is defined by

\begin{align}
    \text{MOTAR}=\max(0, 1 -  \frac{\text{IDS}_\text{r} + \text{FP}_\text{r} + \text{FN}_\text{r} - \text{(1-r)} \times \text{GT}}{\text{r} \times \text{GT}})
\end{align}
where $\text{IDS}_\text{r}$ is the number of identity switches under that recall, $\text{FP}_\text{r}$ is the number of false positives under that recall, and $\text{FN}_\text{r}$ is the number of false negatives under that recall.

Another key metric used by nuScenes is AMOTP. For a given frame, AMOTP is defined by
\begin{equation}
    \text{AMOTP} = \frac{1}{\text{n-1}} \sum_{\text{r} \in \{ \frac{1}{\text{n-1}},  \frac{2}{\text{n-1}}, ... , 1\} } \frac{\sum_i{d_i}}{\sum{\text{TP}}},
\end{equation}
where $d_i$ is the Euclidean distance between the $i^{\text{th}}$ TP detection position and its corresponding ground truth position.

\end{document}